\documentclass[12pt]{article}

\usepackage[authoryear]{natbib}
\usepackage{graphicx}
\usepackage{caption}
\usepackage{subcaption}
\usepackage{amssymb}
\usepackage{amsmath}
\usepackage{times}
\usepackage{bm}
\usepackage{amsmath,amsfonts,amssymb,theorem,latexsym,hyperref,
            epsfig,multicol,multirow,
            anysize,graphicx,enumitem,wasysym,
            tipa,txfonts,pxfonts, mathrsfs}
\usepackage{lscape}
\usepackage{threeparttable}
\usepackage{float}
\usepackage{color}

\newcommand{\underl}{\underline}

\title{Box-Cox symmetric distributions and applications to nutritional data}
\author{Silvia L.P.~Ferrari\thanks{Corresponding author. Email: silviaferrari@usp.br}\\ 
{\small {\em Department of Statistics, University of S\~ao Paulo, S\~ao Paulo, Brazil}}
\\
Giovana Fumes\\
{\small {\em Department of Exact Sciences, ESALQ, University of S\~ao Paulo, Piracicaba, Brazil}}
}

\date{}

\begin{document}
\maketitle

\begin{abstract}
We introduce and study the Box-Cox symmetric class of distributions, which is useful for
modeling positively skewed, possibly heavy-tailed, data. The new class of distributions includes the
Box-Cox t, Box-Cox Cole-Green {\color{black}{(or Box-Cox normal)}}, Box-Cox power exponential distributions, and the class of the log-symmetric distributions as special cases. It provides easy parameter interpretation, which makes it convenient for regression modeling purposes. Additionally, it provides enough flexibility to handle outliers.
The usefulness of the Box-Cox symmetric models is illustrated in a series of applications to nutritional data.\\

\end{abstract}

\section{Introduction}\label{introduction}

It is well-known that positive continuous data usually present positive skewness and outlier observations.
This is the typical situation with survival times, nutrients intake and family income data, among many other examples.
Since \cite{Box} seminal paper, the Box-Cox power transformation has been routinely employed
for transforming to normality. Let $Y$ be a positive random variable. The Box-Cox transformation is defined as
{\color{black}$(Y^\lambda-1)/{\lambda},$ if $\lambda \ne 0$, and $\log Y,$ if $\lambda = 0$.}
Despite its popularity and ease of implementation, this approach, however, has
drawbacks, one of them being the fact that the model parameters cannot be easily interpreted in terms of the
original response. A conceptual shortcoming is that the support of the transformed variable is not the whole real
line and hence a (non-truncated) normal distribution should not be assumed for the transformed data.

An alternative approach to the Box-Cox transformation that allows the parameters to be interpretable as characteristics of the original
data is the Box-Cox Cole-Green distribution {\color{black}{(or Box-Cox normal distribution)}}; see \citet{COLE}. It uses the Box-Cox approach, but the parameters 
are incorporated into the transformation. The Box-Cox Cole-Green distribution has support in $\mathbb{R}^+$ and is defined from the transformation
{\color{black}
\begin{equation}
Z\equiv h(Y; \mu,\sigma,\lambda)=\left\{
\begin{array}{ll}
 {\frac{1}{\sigma \lambda }\left[\left(\frac{Y}{\mu}\right)^{\lambda}-1\right]}, &{\mbox{if} \quad \lambda\neq0}, \\
 {\frac{1}{\sigma}\log\left(\frac{Y}{\mu}\right)}, &{ \mbox{if} \quad \lambda= 0,}
\end{array}
\label{ztrans}
\right.
\end{equation}
}
where $\mu>0$, $\sigma>0$, $-\infty<{\color{black}{\lambda}}<\infty$, assuming that $Z$ has a standard normal distribution truncated at a suitable interval of the real line;
details are given in the next section.
The Box-Cox symmetric (BCS) class of distributions defined in this paper replaces the normal distribution by the class of the continuous standard symmetric
distributions. Replacing the normal distribution by the Student-t and the power exponential distributions results in the Box-Cox t \citep{Rigby2} and the
Box-Cox power exponential \citep{rigby12, Voudouris} distributions. Additionally, it generalizes the class of the log symmetric distributions \citep{Vanegas1, Vanegas}.

The paper unfolds as follows. In Section \ref{BCS}, the Box-Cox symmetric class of distributions is defined, some properties are stated and
interpretation of the parameters in terms of quantiles is discussed. Tail heaviness of Box-Cox symmetric distributions is studied in Section \ref{Tail}. It is shown that
the Box-Cox symmetric class of distributions allows much more tail flexibility than the log-symmetric distributions.
{\color{black}{Likelihood-based inference in discussed in Section \ref{inference}.}} It is suggested that the choice of the symmetric distribution may lead to robust
estimation against outliers. In Section \ref{applic}, applications to 33 nutrients intake data are presented, and a comparison of alternative approaches is provided.
Finally, concluding remarks (Section \ref{conc}) close the paper. {\color{black}{Technical details are left for the Appendix.}}

\section{Box-Cox symmetric distributions}\label{BCS}

A continuous random variable $W$ is said to have a symmetric distribution with location  parameter ${\color{black}{\mu}} \in \mathbb{R}$, scale parameter
${\color{black}{\sigma}}>0$ and density generating function $r$, and we write $W \sim S({\color{black}{\mu}},{\color{black}{\sigma}}; r)$, if its probability density function (pdf) is given by
\begin{eqnarray}\label{sim}
{\color{black}{
f_W(w)=\frac{1}{{\color{black}{\sigma}}} r\left( \left(\frac{w-{\color{black}{\mu}}}{{\color{black}{\sigma}}}\right)^{2}\right),}} \qquad w \in \mathbb{R},
\end{eqnarray}
where $r(\cdot)$ satisfies $r(u)>0$, for $u\ge 0$, and $\int_0^{\infty} u^{-1/2} r(u) {\rm d}u =1$.
The class of the symmetric distributions has a number of well-known distributions
as special cases depending on the choice of $r$. It includes the normal distribution as well as the Student-t, power exponential, type I logistic,
type II logistic and slash distributions among others. Densities in this family have quite different tail behaviors, and some of them may have heavier or lighter tails than the normal distribution.

The symmetric distributions have some interesting properties. Some of them follow:
{\it (i)}
If $W \sim S({\color{black}{\mu}},{\color{black}{\sigma}}; r)$, its characteristic function is $\psi_W(t)= e^{it{\color{black}{\mu}}} \varphi{(t^2 {\color{black}{\sigma}}^2)}$, $t \in \mathbb{R}$, for some
function $\varphi$, with $\varphi(u) \in \mathbb{R}$, for $u>0$. Whenever they exist, ${\rm E}(W)={\color{black}{\mu}}$ and ${\rm Var}(W)=\xi {\color{black}{\sigma}}^2$, where
$\xi=-2 \varphi'(0)>0$, with $\varphi'(0)={\rm d} \varphi(u)/{\rm d} u |_{u=0}$ is a constant not depending on ${\color{black}{\mu}}$ and ${\color{black}{\sigma}}$ \citep{Fang}.
If $u^{-(k+1)/2} r(u)$ is integrable, then the $k$th moment of $W$ exist \citep{kelker}.
{\it (ii)}
If $W \sim S({\color{black}{\mu}},{\color{black}{\sigma}}; r)$, then $a+bW \sim S(a+b{\color{black}{\mu}},|b| {\color{black}{\sigma}}; r)$, where $a,b \in \mathbb{R}$, with $b \neq 0$.
In particular, if $W \sim S({\color{black}{\mu}},{\color{black}{\sigma}}; r)$, then $S=(W-{\color{black}{\mu}})/{\color{black}{\sigma}} \sim S(0,1; r)$, {\color{black}{and its pdf is 
$f_S(s)=r(s^2),$ for $s \in \mathbb{R}$.}}

{\color{black}{
Let $Y$ be a positive continuous random variable, and consider $Z\equiv h(Y;\mu,\sigma,\lambda)$ as in (\ref{ztrans}). Assume that $Z$ has a standard symmetric 
distribution truncated at $\mathbb{R}\backslash A(\sigma,\lambda)$, where
\begin{equation}
A(\sigma, {\color{black}{\lambda}})=\left\{
\begin{array}{ll}
 {\left(-\frac{1}{\sigma {\color{black}{\lambda}}},  \infty \right)}, &{ \mbox{if} \quad {\color{black}{\lambda}}>0}, \\
 {\left(-\infty,-\frac{1}{\sigma {\color{black}{\lambda}}}\right)}, &{ \mbox{if} \quad {\color{black}{\lambda}}<0,} \\
 {\left(-\infty, \infty \right)}, &{ \mbox{if} \quad {\color{black}{\lambda}}=0,}
\end{array}
\label{ind}
\right.
\end{equation}
i.e. the support of the truncated distribution is $A(\sigma,\lambda)$. We then say that $Y$ has a Box-Cox symmetric distribution with parameters $\mu>0$, $\sigma>0$, and $\lambda \in \mathbb{R}$, and density generating function $r$,
and we write $Y \sim BCS(\mu, \sigma, {\color{black}{\lambda}}; r)$. In other words, $Y \sim BCS(\mu, \sigma, {\color{black}{\lambda}}; r)$ if the transformed variable $Z$ in (\ref{ztrans}) has the distribution of $S \sim S(0,1;r)$ truncated at $\mathbb{R}\backslash A(\sigma,\lambda)$. 
}}

The Box-Cox symmetric class of distributions reduces to the log-symmetric class of distributions \citep{Vanegas1} when ${\color{black}{\lambda}}$ is fixed at zero.
Additionally, it leads to the Box-Cox Cole-Green \citep{stasinopoulos2008}, Box-Cox t \citep{Rigby2} and Box-Cox power exponential \citep{rigby12, Voudouris} distributions by taking $Z$ as a truncated standard normal, Student-t and power exponential random variable, respectively. The density generating function, $r(u)$, for $u\ge 0$, for various distributions in the BCS class follows:
\begin{enumerate}[label=(\roman{*}), ref=(\roman{*})]
\item normal: $r(u)= (2\pi)^{-1/2} \exp \{-u/2 \}$;
\item double exponential: $r(u)=(\sqrt{2}/2)\exp\{-\sqrt{2}u^{1/2}\}$;
\item power exponential: $r(u)=[\tau/(p(\tau) 2^{1+1/\tau} \Gamma(1/\tau))] \exp\{-u^{\tau/2}/(2 p(\tau)^{\tau})\}]$, where $\tau>0$ and $p(\tau)^2= 2^{-2/\tau} \Gamma(1/\tau) [\Gamma(3/\tau)]^{-1}$; when $\tau=1$ and $\tau=2$, $r(u)$ coincides with the density generating function of the double exponential and normal, respectively;
\item Cauchy: $r(u)=\{\pi (1+u)\}^{-1}$;
\item Student-t: $r(u)=\tau^{\tau/2} \{B(1/2,\tau/2)\}^{-1}(\tau+u)^{-(\tau+1)/2}$, $\tau>0$, where $B(\cdot,\cdot)$ is the beta function;
\item type I logistic: $r(u)= c \exp \{-u \}(1+\exp \{-u \})^{-2}$, where $c \approx 1.484300029$ is the normalizing constant,
obtained from the relation $\int_{0}^{\infty} u^{-1/2} r(u){\rm d}u=1$;
\item type II logistic: $r(u)= \exp \{-u^{1/2}\} (1+\exp \{-u^{1/2}\})^{-2}$;
\item canonical slash \citep{Rogers}: $r(u)= (1/\sqrt{2\pi}u) (1-\exp\{-u/2\})$, for $u > 0$, and $r(u)=1/(2 \sqrt{2\pi})$,  for $u = 0$;
\item slash\footnote{It is the distribution of $Z/U^{1/q}$, where $q>0$ and $Z$ and $U$ are independent random variables with standard normal and uniform distribution, respectively.}:
$r(u)= \Psi((q+1)/2,u/2) q 2^{q/2-1}/(\sqrt{\pi} u^{(q+1)/2})$, for $u>0$, and $r(u)= q/[(q+1)\sqrt{2\pi}]$, for $u=0$, $q>0$, where $\Psi(a,x)=\int_{0}^{x} t^{a-1} e^{-t} {\rm d}t$ is the lower incomplete gamma function; when $q=1$ the slash distribution coincides with the canonical slash distribution.
\end{enumerate}

{\color{black}{
Let $f_{Z}(\cdot)$ be the pdf of $Z$ with $Z$ given in (\ref{ztrans}). We have
\begin{eqnarray}\label{densss}
f_{Z}(z)=\frac{f_{S}(z)}{F_{S}\left(\frac{1}{\sigma \mid {\color{black}{\lambda}} \mid}\right)}, \quad z \in A(\sigma, {\color{black}{\lambda}}),
\end{eqnarray}
where  $f_S(\cdot)$ and $F_{S}(\cdot)$ are the pdf and the cumulative distribution functions (cdf) of $S \sim S(0,1; r)$,
respectively.}}\footnote{If $\sigma |{\color{black}{\lambda}}|=0$, $1/\sigma|{\color{black}{\lambda}}|$ is interpreted as $\lim_{\sigma {\color{black}{\lambda}} \rightarrow 0}{( 1/\sigma |{\color{black}{\lambda}}|
 )}=\infty$ and $F ( 1/\sigma |{\color{black}{\lambda}}|)$ is taken as 1.}
{\color{black}{ Now, let $z={\color{black}{h}}(y;\mu,\sigma,{\color{black}{\lambda}})$ (see (\ref{ztrans})).
Because the Jacobian of the transformation from $y$ to $z$ is 
$|{\partial z}/{\partial y}| =  y^{\lambda-1}/{\mu^\lambda}\sigma$
the pdf of $Y$ is given by
\begin{eqnarray} \label{dens111}
f_{Y}(y)=\frac{y^{\lambda-1}}{\mu^\lambda\sigma}f_{Z}(z), \quad y>0.
\end{eqnarray}
}}
Since $f_S(s)=r(s^2)$, we have from (\ref{ztrans}) and (\ref{densss}) that (\ref{dens111}) can be written as
\begin{equation}
f_Y(y)= \left\{
\begin{array}{ll} \label{fgdens}
{\frac{y^{{\color{black}{\lambda}}-1}}{\mu^{\color{black}{\lambda}} \sigma} \frac{r\left(z^2 \right)}{R \left(\frac{1}{\sigma |{\color{black}{\lambda}}|}\right)} },&{\mbox{if} \quad {\color{black}{\lambda}}\neq0},  \\
{\frac{1}{y \sigma} r\left( z^2 \right)}, &{\mbox{if} \quad {\color{black}{\lambda}}= 0,}
\end{array}
\right.
\end{equation}
for $y>0$, where $R({\color{black}{s}})=\int_{-\infty}^{{\color{black}{s}}} r(u^2) {\rm d}u$, for ${\color{black}{s}} \in \mathbb{R}$.
The cdf of $Y$ is given by
$$
F_{Y}(y)= \left\{
\begin{array}{ll}
{\frac{R(z)}{R \left(\frac{1}{\sigma \mid {\color{black}{\lambda}} \mid}\right)}}, &{\mbox{if} \quad {\color{black}{\lambda}} \leq 0},\\
{\frac{R(z)-R\left(-\frac{1}{\sigma \mid {\color{black}{\lambda}} \mid}\right)}{R \left(\frac{1}{\sigma \mid {\color{black}{\lambda}} \mid}\right)}},
 &{\mbox{if} \quad {\color{black}{\lambda}}>0},
\end{array}
\label{acumulada}
\right.
$$
for $y>0$.

\begin{figure}[ht]
\centering
\includegraphics[width=0.35\hsize]{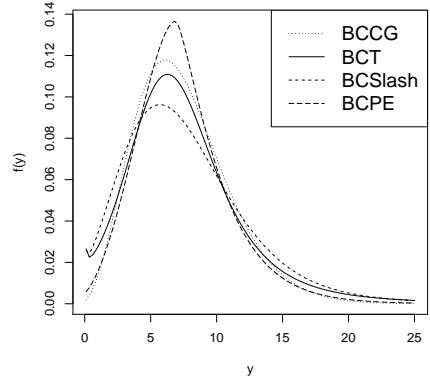}
\caption{Probability density functions of BCCG, BCT ($\tau=4$), BCPE ($\tau=1.5$) and BCSlash ($q=4$) for $\mu=7$, $\sigma=0.5$ and ${\color{black}{\lambda}}=0.5$.}
\label{comparandoBCS}
\end{figure}

\begin{figure}[ht]
\centering
   \begin{subfigure}[b]{0.35\textwidth}
    \includegraphics[width=1 \hsize]{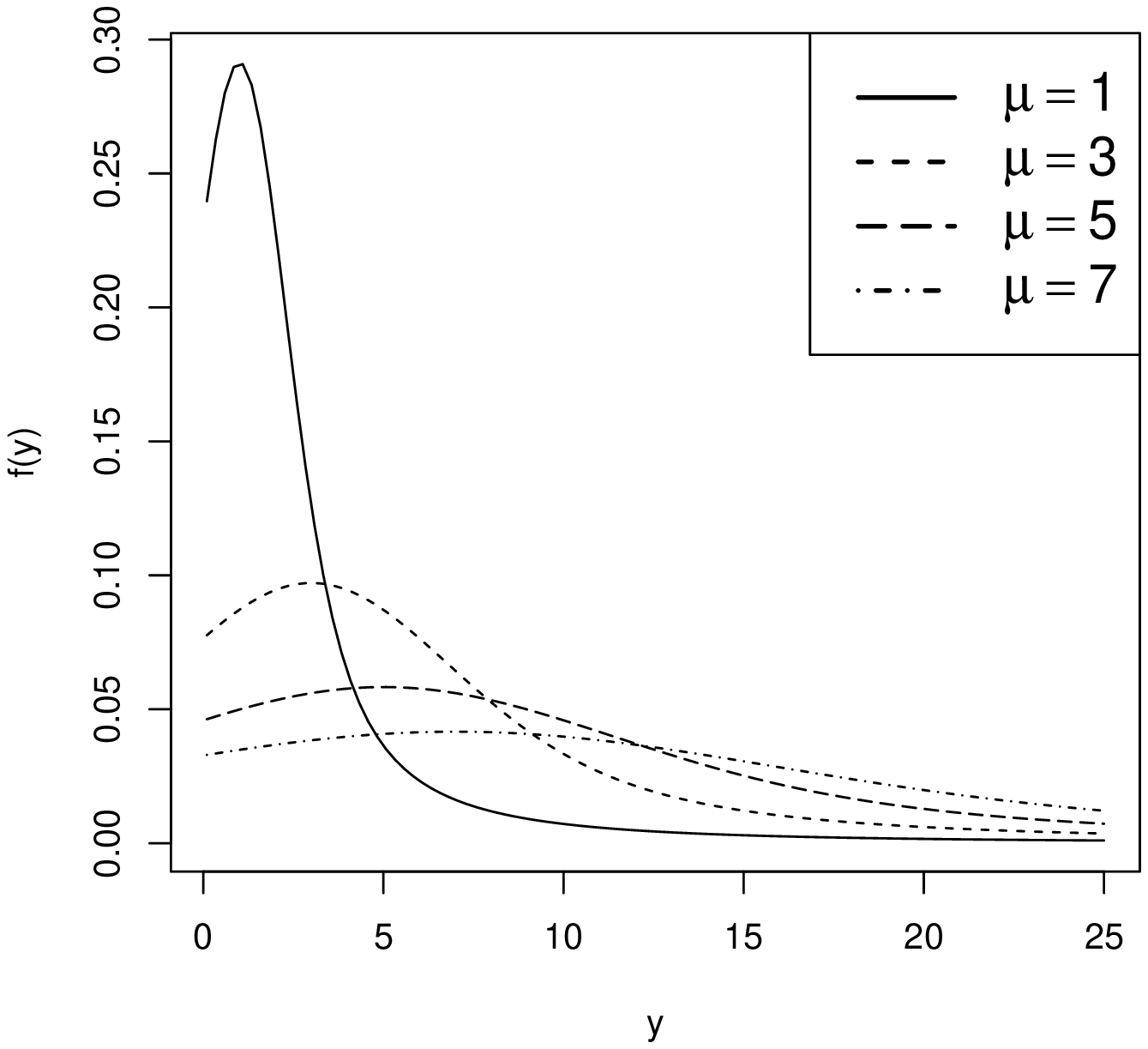}
                \caption{$\sigma=1$; ${\color{black}{\lambda}}=1$; $q=1$.}
                \label{medianaslash}
   \end{subfigure}
   \begin{subfigure}[b]{0.35\textwidth}
               \includegraphics[width=1 \hsize]{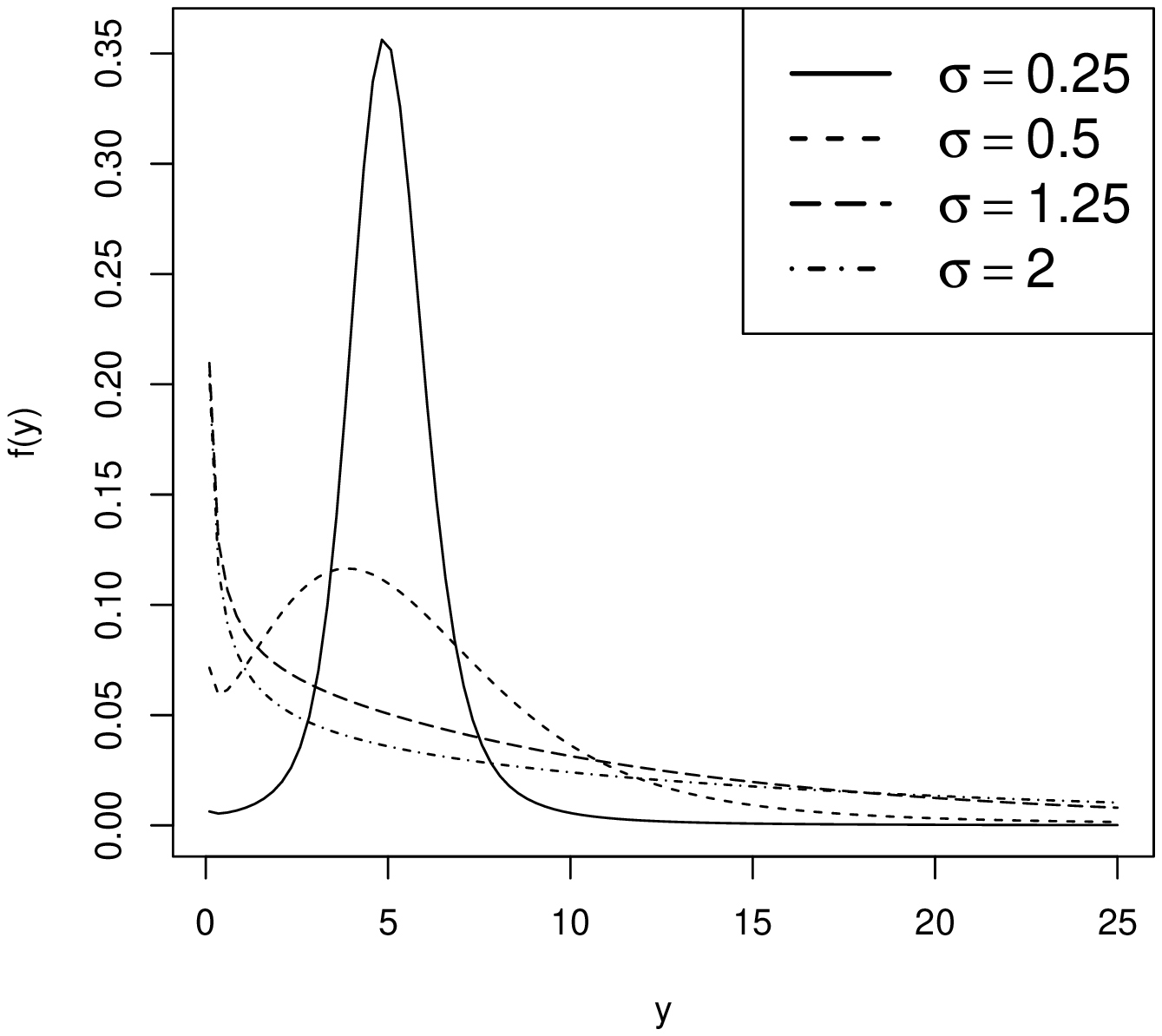}
                \caption{$\mu=5$; ${\color{black}{\lambda}}=0.5$; $q=2$.}
                \label{coeficientedevariacaoslash}
   \end{subfigure}
   \begin{subfigure}[b]{0.35\textwidth}
              \includegraphics[width=1 \hsize]{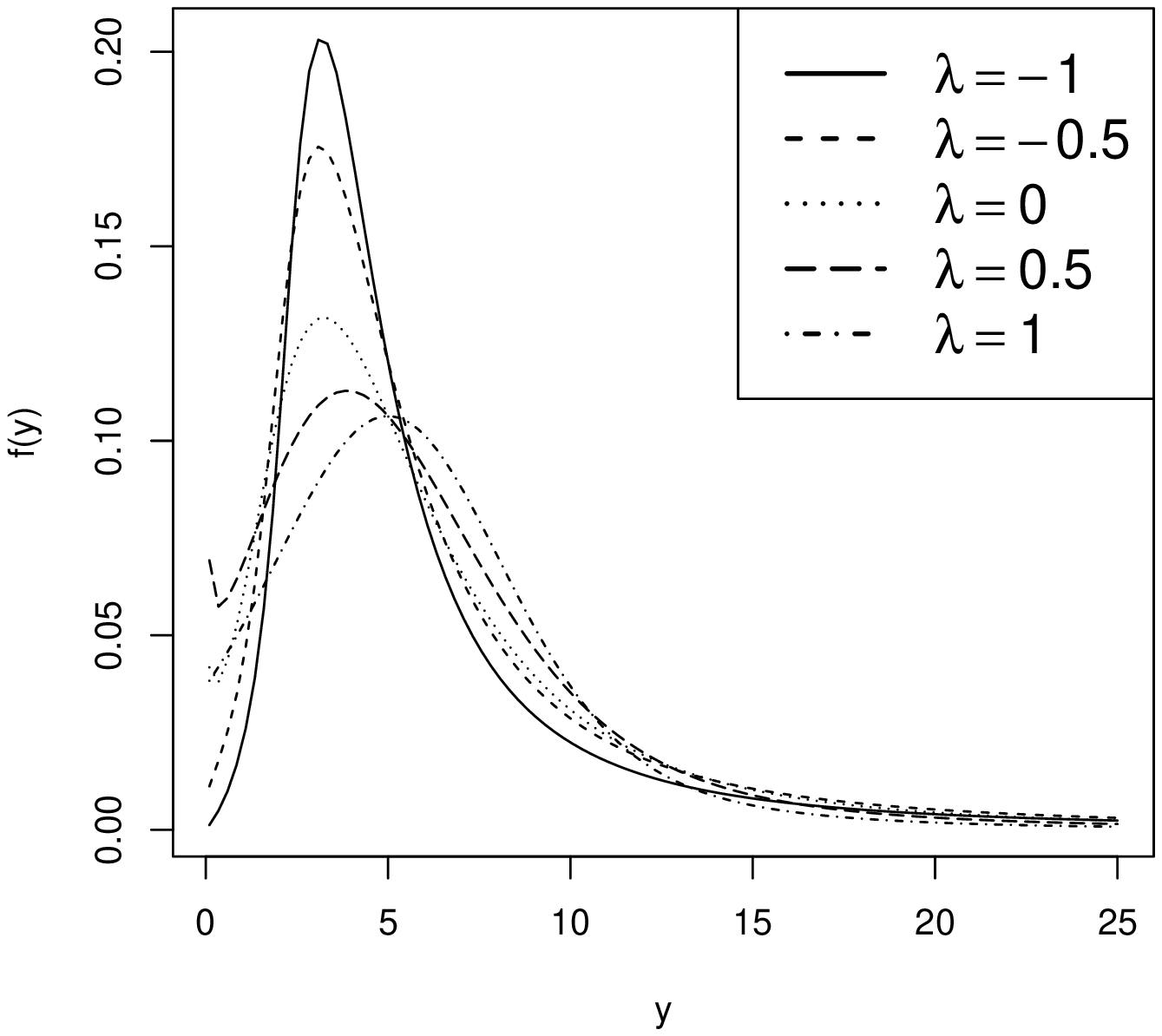}
                \caption{$\mu=5$; $\sigma=0.5$; $q=2$.}
                \label{transformacaoslash}
   \end{subfigure}
   \begin{subfigure}[b]{0.35\textwidth}
             \includegraphics[width=1 \hsize]{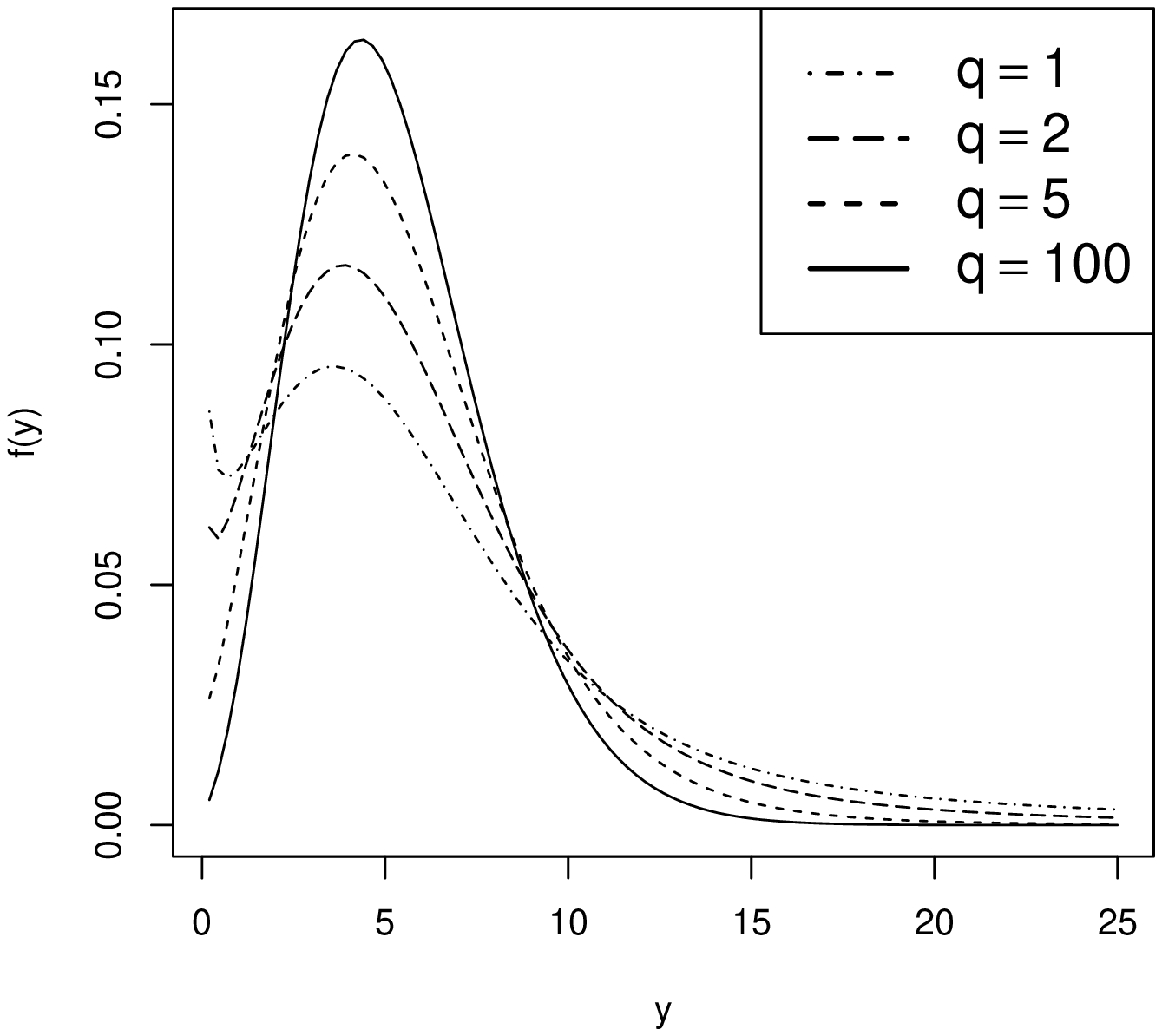}
                \caption{$\mu=5$; $\sigma=0.5$; ${\color{black}{\lambda}}=0.5$.}
                \label{qslash}
   \end{subfigure}
\caption{Probability density functions of $BCSlash(\mu,\sigma,{\color{black}{\lambda}},q)$. }
\label{slash}
\end{figure}

\cite{Rigby2} and \cite{Voudouris} present figures of probability density functions of the Box-Cox t and the Box-Cox power exponential distributions for different values of the parameters. The figures suggest, as expected, that the transformation parameter ${\color{black}{\lambda}}$ controls the skewness of the distribution, while the right tail/kurtosis behavior is controlled by the extra parameter (degrees of freedom parameter of the Box-Cox t distribution and the shape parameter of the Box-Cox power exponential distribution). Figure \ref{comparandoBCS} shows the pdf of the Box-Cox Cole-Green (BCCG), Box-Cox t (BCT), Box-Cox power exponential (BCPE) and Box-Cox slash (BCSlash) distributions for a particular choice of the parameters. It is apparent that the BCT and BCSlash distributions have heavier right tail than the other distributions. Figure \ref{slash} shows the pdf of the BCSlash distribution for different values of the parameters. Note that the extra parameter $q$ controls for right tail heaviness; see Section \ref{Tail} for a detailed discussion of tail
heaviness of the Box-Cox symmetric distributions.

It is straightforward to verify that if $Y \sim BCS(\mu, \sigma, {\color{black}{\lambda}}; r)$, for $\mu>0$, $\sigma>0$ and ${\color{black}{\lambda}} \in \mathbb{R}$, the following properties hold:
\begin{list}{}{\setlength{\rightmargin0in}{\leftmargin0.3in}}
\item[(i)] $dY \sim BCS (d \mu, \sigma, {\color{black}{\lambda}}; r)$, for all constant $d>0$, and hence $\mu$ is a scale parameter;

\item [(ii)] $({Y}/{\mu})^d \sim BCS (1, d\sigma, {\color{black}{\lambda}}/d; r)$, for all constant $d>0$. In particular, ${Y}/{\mu} \sim BCS(1,\sigma,{\color{black}{\lambda}}; r)$, $({Y}/{\mu})^{{1}/{\sigma}} \sim BCS(1,1, \sigma {\color{black}{\lambda}}; r)$, and, for ${\color{black}{\lambda}} > 0$, $({Y}/{\mu})^{{\color{black}{\lambda}}} \sim BCS(1, {\color{black}{\lambda}}\sigma, 1; r)$;

\item[(iii)] if ${\color{black}{\lambda}}=1$ then $Y$ has a truncated symmetric distribution with parameters $\mu$ and $\mu\sigma$ and support
$(0,\infty)$;

\item[(iv)] from (\ref{fgdens}), we have that, if ${\color{black}{\lambda}}=0$, $Y$ has
a log-symmetric distribution with parameters $\mu$ and $\sigma^2$ and density generating function $r$ \citep{Vanegas1};
it is denoted here by $\emph{LS}(\mu, \sigma; r)$.

\item[(v)] for integer $k$,
\begin{equation}
{\rm E}(Y^k)=
\left\{
\begin{array}{ll}
\mu^k {\rm E} \left( (1+ \sigma {\color{black}{\lambda}} \emph{S})^\frac{k}{{\color{black}{\lambda}}} \emph{I}_{\emph{A}(\sigma,{\color{black}{\lambda}})}(\emph{S}) \right),
&{\mbox{if} \quad {\color{black}{\lambda}} \neq 0,}  \nonumber \\
\mu^k {\rm E} \left(\exp(k \sigma S)\right),&{\mbox{if} \quad {\color{black}{\lambda}} = 0,} \nonumber
\end{array}
\right.\label{momentos}
\end{equation}
where $I_{A(\sigma,{\color{black}{\lambda}})}(\cdot)$ is the indicator function of the set $A(\sigma,{\color{black}{\lambda}})$ given in (\ref{ind}), and $S\sim S(0,1;r)$.
Hence, when ${\color{black}{\lambda}}=0$ (or the truncation set $\mathbb{R}\backslash A(\sigma,{\color{black}{\lambda}})$ has negligible probability under the $S(0,1;r)$ distribution) the moments of $Y$ can be obtained from the characteristic function of a standard symmetric distribution.
\end{list}

The BCS class of distributions allows easy parameter interpretation from its quantiles. Let $s_\alpha$ denote the $\alpha$-quantile of the 
${\color{black}{S \sim }}S(0,1;r)$, and $z_\alpha$ be the $\alpha$-quantile of {\color{black}{the truncated $S$, i.e. of the standard symmetric distribution $S(0,1;r)$ truncated at $\mathbb{R}\backslash A(\sigma,{\color{black}{\lambda}})$}}. We have
$$
z_{\alpha}= \left\{
\begin{array}{ll}
{R^{-1} \left[ \alpha R\left(\frac{1}{\sigma\mid{\color{black}{\lambda}}\mid}\right)\right]}, {\quad \mbox{if} \quad {\color{black}{\lambda}} \leq 0}, \\
{R^{-1} \left[ 1-(1-\alpha)R\left(\frac{1}{\sigma\mid{\color{black}{\lambda}}\mid}\right)\right]}, {\quad \mbox{if} \quad {\color{black}{\lambda}} > 0}.
\end{array}
\right.\label{quantis}
$$
Recall that $R(\cdot)$ is the cdf of $S \sim S(0,1; r)$.
If $Y \sim BCS(\mu, \sigma, {\color{black}{\lambda}}; r)$ its $\alpha$-quantile is given by
\begin{equation}
y_{\alpha}= \left\{
\begin{array}{ll}
{\mu(1+ \sigma {\color{black}{\lambda}} z_{\alpha})^{\frac{1}{{\color{black}{\lambda}}}}}, &{\mbox{if} \quad {\color{black}{\lambda}} \neq 0,} \nonumber \\
{\mu \exp(\sigma z_{\alpha})}, &{\mbox{if} \quad {\color{black}{\lambda}}=0}.\nonumber
\end{array}
\right.
\end{equation}

We note that all the quantiles are proportional to $\mu$. In particular, if ${\color{black}{\lambda}}=0$ we have $z_{1/2}=0$ and hence $\mu$ is the median of $Y$. Moreover,
if the truncation set $\mathbb{R}\backslash A(\sigma,{\color{black}{\lambda}})$ has negligible probability under the $S(0,1;r)$ distribution {\color{black}{(this happens when $\sigma\lambda$ is small)}}, $\mu$ is approximately equal to the median of $Y$. In order to give an interpretation for $\sigma$, we consider the centile-based coefficient of variation \citep{Rigby2}
$$CV_Y=\frac{3}{4}\frac{y_{0.75}-y_{0.25}}{y_{0.5}}.$$ When ${\color{black}{\lambda}}$ is not far from zero and ignoring the truncation region, $CV_Y\approx 1.5\sinh(\sigma s_{0.75})$, which is an increasing function of $\sigma$. Here, $\sinh(\cdot)$ is the hyperbolic sine function. The approximation is exact when ${\color{black}{\lambda}}=0$. Therefore, $\sigma$ can be seen as a relative dispersion parameter. Finally, ${\color{black}{\lambda}}$ is regarded as a skewness parameter since it determines the power transformation to symmetry.

At this point it is informative to compare our approach with an alternative, closely related, approach
{\color{black}{that uses the original Box-Cox transformation}}.
A usual strategy for dealing with positive continuous asymmetric data is to employ a Box-Cox transformation in the data 
and to assume that the transformed data follow a normal distribution.
The normal distribution can be replaced by the class of the continuous symmetric distributions in the Box-Cox transformation approach; see \cite{gauss}. Formally, this approach does not correspond to assume a coherent distribution for the data because the support of the transformed variable is not the entire real line, unless the transformation parameter is zero{\color{black}{; this is known as the truncation problem. In order to take the correct support of the Box-Cox transformed data into account, a truncated normal (or symmetric) distribution may be assumed for the transformed data; see e.g. \citet{Poirer} and \citet{Yang96}. However, the truncation point will depend on the three parameters, namely the location, dispersion and transformation parameters. In our approach the truncation point depends on $\sigma$ and $\lambda$ only. Hence, it does not depend on the regressor values if a regression model is assumed for $\mu$. 
Although the truncation problem is usually disregarded in the statistical literature, alternative transformations have been proposed to overcome this problem; see, e.g., \citet{YeoJohnson} and \citet{Yang2006}.
}} 
Furthermore, the model parameters are interpreted as characteristics of the transformed data, not the original data. Our approach does not have these two shortcomings: a genuine distribution is assumed for the data and the parameters are interpretable in terms of characteristics of the original data, not the transformed data. 

In Section \ref{applic}, we present a comparison of alternative approaches in real data applications.

\section{Tail heaviness}\label{Tail}

Heavy-tailed distributions have been frequently used to model phenomena in various fields such as finance and environmental sciences;
see, for instance, \cite{resnick}. A usual criterion for evaluating tail heaviness in the extreme value theory
is the tail index of regular variation functions. Informally, a regular variation function behaves asymptotically as
a power function. Formally, a Lebesgue measurable function $M: \mathbb{R}^+ \rightarrow \mathbb{R}^+$ is regularly varying at infinity
with index of regular variation $\alpha$ ($M \in RV_\alpha$ ), if $\lim_{t \rightarrow \infty}M(ty) / M(t)=y^{\alpha}$ for $y>0$.
If $\alpha=0$, $M$ is said to be a slowly varying function.
The function $M$ varies rapidly at infinity or is regularly varying at infinity with index $-\infty$ ($\infty$), or
$M \in RV_{-\infty}$ ($M \in RV_\infty$), if, for $y>0$,
$\lim_{t\rightarrow \infty}{M(ty)}/{M(t)}:=y^{-\infty}$ ($\lim_{t\rightarrow \infty}{M(ty)}/{M(t)}:=y^{\infty}$); see \citet[p. 4]{haan}.
\footnote{$y^{-\infty}=\infty$, if $0<y<1$, $=1$, if $ y=1$, $=0$, if $y>1$; \ $y^{\infty}=0$, if $0<y<1$, $=1$, if $ y=1$, $=\infty$, if $y>1$.}

A continuous distribution with cdf $F$ is said to have a heavy right tail whenever  $\overline{F}:=1-F$
is a regularly varying at infinity function with negative index of regular variation $\alpha=-1/\xi, \ \ \xi>0$, i.e.,
$\lim_{t \rightarrow \infty} \overline{F}(ty) / \overline{F}(t)=y^{-1/\xi}$.
The parameter $\xi$ is called the tail index. From the l'H\^opital rule,
this limit can be written as $y\lim_{t\rightarrow \infty} f(ty)/f(t)$, for $y>0$, where $f$ is the pdf corresponding to $F$.
When the limit equals $y^{-\infty}$ we say that $F$ has a light (non-heavy) right tail and that the tail index is zero.
When the limit equals 1, i.e. $\overline F$ is a slowly varying function, we will say that $F$ has right heavy tail with tail index $\infty$.

From \citet[Corollary 1.2.1]{haan} it follows that the tail index is invariant to location-scale transformations.
Hence, from
(\ref{sim}) the tail index of a $S(\mu,\sigma;r)$ distribution is independent of $\mu$ and $\sigma$ and is obtained from
\begin{equation}\label{icsim}
\mathcal{L}_{S}(w;r)= w \lim_{t \rightarrow \infty} \frac{r(t^2w^2)}{r(t^2)}.
\end{equation}
It can be shown that the tail index of the $BCS(\mu,\sigma,{\color{black}{\lambda}};r)$ distribution, for all $\mu >0$ and all $\sigma>0$, can be obtained from
$$
\mathcal{L}_{BCS}(y, \mu, \sigma, {\color{black}{\lambda}}; r)=\left\{
\begin{array}{ll}
\displaystyle {\mathcal{L}_{S}(y^{\color{black}{\lambda}}; r)},{\quad \mbox{if} \quad {\color{black}{\lambda}}>0,} \\
\displaystyle {\mathcal{L}_{LS}(y, \mu, \sigma; r)}, {\quad \mbox{if} \quad {\color{black}{\lambda}}=0,}\\
\displaystyle {y^{\color{black}{\lambda}}}, {\quad \mbox{if} \quad {\color{black}{\lambda}}<0,}\\
\end{array}
\label{icbcs}
\right.
$$
where $\mathcal{L}_{LS}(y, \mu, \sigma; r)$ corresponds to the limit that defines the tail index of the log-symmetric distributions
and is given by (\ref{icsim}) with $tw$ in the numerator replaced by $\sigma^{-1}\log(tw/\mu)$ and $t$ in the denominator replaced by
$\sigma^{-1}\log(t/\mu)$.

Table \ref{index} gives the tail index of some Box-Cox symmetric distributions.\footnote{The tail indices were obtained using Maple 13; see http://www.maplesoft.com. The tail index for the log-power exponential distribution with $\tau>1$ was obtained for $\tau \in \mathbb{Q}$, and for the slash distribution, for $q \in \mathbb{N}^{*}$.}
When ${\color{black}{\lambda}}>0$, the Box-Cox t and Box-Cox slash distributions have heavy right tail with the extra parameter
(the degrees of freedom parameter $\tau$ and the shape parameter $q$ in the case of the t and the slash distributions, respectively) controlling
the tail weight for fixed ${\color{black}{\lambda}}$; the Box-Cox Cole-Green (normal), Box-Cox power exponential, Box-Cox type I logistic and Box-Cox type II logistic distributions are all right light-tailed distributions, i.e. the tail index for these distributions are zero.
The results for ${\color{black}{\lambda}}=0$ reveal that the log-normal, log-power exponential with $\tau>1$ and log-type I logistic distributions are right light-tailed while the log-double exponential and the log-type II logistic distributions have heavy right tail with tail index determined by $\sigma$. All the others have right heavy tail with tail index $\infty$. It is noteworthy that the extra parameters have no influence on the tail index. This suggests that the class of the log-symmetric distributions \citep{Vanegas1} is much more restrictive than the Box-Cox symmetric distributions defined in this paper with respect to tail flexibility.
Finally, when ${\color{black}{\lambda}}<0$, all the Box-Cox symmetric distributions have heavy right tail with tail index equal to $|{\color{black}{\lambda}}|^{-1}$.

\begin{table}[h]
\normalsize \caption{Tail index ($\xi$) of some symmetric and Box-Cox symmetric distributions.}
\vspace{-0.8cm}
\begin{center}
\begin{tabular}{l|l|l|l|l}
\hline
distribution                   & symmetric&BCS (${\color{black}{\lambda}}>0$) &BCS (${\color{black}{\lambda}}=0$)&BCS (${\color{black}{\lambda}}<0$) \\
\hline
normal                         & $0$       & $0$             & $0$               & $1/|{\color{black}{\lambda}}|$   \\
double exponential             & $0$       & $0$             & $\sigma/\sqrt{2}$ & $1/|{\color{black}{\lambda}}|$   \\
power exponential              &           &                 &                   &         \\
\hspace{2cm}  $\tau>1$         & $0$       & $0$             & $0$               & $1/|{\color{black}{\lambda}}|$       \\
\hspace{2cm}  $\tau=1$         & $0$       & $0$             & $\sigma/\sqrt{2}$ & $1/|{\color{black}{\lambda}}|$       \\
\hspace{2cm}  $\tau<1$         & $0$       & $0$             & $\infty$          & $1/|{\color{black}{\lambda}}|$       \\
Cauchy                         & 1         & $1/{\color{black}{\lambda}}$         & $\infty$          & $1/|{\color{black}{\lambda}}|$ \\
t                              & $1/\tau$  & $1/({\color{black}{\lambda}} \tau)$  & $\infty$          & $1/|{\color{black}{\lambda}}|$ \\
type I logistic                & $0$       & $0$             & $0$               & $1/|{\color{black}{\lambda}}|$   \\
type II logistic               & $0$       & $0$             & $\sigma$          & $1/|{\color{black}{\lambda}}|$   \\
canonical slash                & 1         & $1/{\color{black}{\lambda}}$         & $\infty$          & $1/|{\color{black}{\lambda}}|$ \\
slash ($q \in \mathbb{N}^{*}$) & $1/q$     & $1/({\color{black}{\lambda}} q)$     & $\infty$          & $1/|{\color{black}{\lambda}}|$   \\
\hline
\end{tabular}
\end{center}
\label{index}
\end{table}

An alternative approach to compare tail heaviness of statistical distributions is considered by \citet[Chapter 12]{Rigby4}.
Here, we focus on right tail heaviness only. If the random variables $Y_1$ and $Y_2$ have continuous pdf's $f_{Y_1}(y)$ and $f_{Y_2}(y)$
and $\lim_{y\rightarrow \infty}f_{Y_1}(y) = \lim_{y\rightarrow \infty}f_{Y_2}(y)= 0$ then $Y_2$ has heavier right tail than $Y_1$
if and only if $\lim_{y\rightarrow \infty}(\log f_{Y_2}(y) - \log f_{Y_1}(y)) =\infty.$
The authors classify the possible asymptotic (large $y$) behavior of the logarithm of a pdf in three major forms:
$-k_2(\log y)^{k_1}$, $-k_4 y^{k_3}$  or $-k_6 \exp(k_5 y)$, with positive $k$'s.
The three forms are decreasing in order of tail heaviness and, for the first form, decreasing $k_1$ results in a heavier tail while decreasing $k_2$ for fixed $k_1$ results in heavier tail. Similarly, for the two other forms.

Table \ref{tail} gives the coefficients of the right tail asymptotic form of the logarithm of pdf's for some symmetric and Box-Cox symmetric distributions.
Following \citet{Rigby4}, the right tail heaviness of the distributions in Table \ref{tail} can be classified in the following four types (the corresponding tail index for each type is given in parentheses):
\begin{itemize}
\item[-] non-heavy tail: type II with $k_3 \geq 1$ ($\xi=0$);
\item[-] heavy tail (i.e. heavier than any exponential distribution) but lighter than any `Paretian type' tail: type I with $k_1 > 1$ and type II with $0 < k_3 < 1$ ($\xi=0$);
\item[-] `Paretian type' tail: type I with $k_1 = 1$ and $k_2 > 1$ ($\xi=1/(k_2-1)$);
\item[-] heavier than any `Paretian type' tail: type I with $k_1 = 1$ and $k_2 = 1$ ($\xi=\infty$).
\end{itemize}
It should be noted that distributions with right tail index $\xi=0$, which are classified as having light
(non-heavy) right tail according to the regular variation theory, are split into two categories in Rigby's criterion:
non-heavy right tail and heavy right tail but lighter than any `Paretian type' tail.

When ${\color{black}{\lambda}}>0$, the Box-Cox t and Box-Cox slash distributions have `Paretian type' right tail with the extra parameter
controlling the right tail heaviness for fixed ${\color{black}{\lambda}}$;
the Box-Cox power exponential distributions have non-heavy right tail ($\tau\ge 1/{\color{black}{\lambda}}$) or heavy right tail but lighter than any `Paretian type' tail
($\tau<1/{\color{black}{\lambda}}$) , with $\tau$ determining the tail heaviness for fixed ${\color{black}{\lambda}}$.
Depending on the value of ${\color{black}{\lambda}}$, the Box-Cox Cole-Green and Box-Cox type I logistic distributions may have a non-heavy right tail (${\color{black}{\lambda}}\ge 1/2$) or a heavy right tail but lighter than any `Paretian type' tail ($0<{\color{black}{\lambda}}<1/2$); similarly for the Box-Cox type II logistic distribution
with ${\color{black}{\lambda}}\ge 1$ and $0<{\color{black}{\lambda}}<1$, respectively.
From the coefficients for ${\color{black}{\lambda}}=0$ we note that, as expected, the log-normal, log-power exponential with $\tau>1$ and log-type I logistic distributions have non-heavy right tail while the log-double exponential and the log-type II logistic distributions have heavy right tail but lighter than any `Paretian type' tail; all the others have right heavier than any `Paretian type' tail.
When ${\color{black}{\lambda}}<0$, all the Box-Cox symmetric distributions in Table \ref{tail} have right `Paretian type' tail with the tail heaviness controlled by
${\color{black}{\lambda}}$ only.

\begin{table}[h]
\scriptsize\normalsize \caption{Coefficients of the right tail asymptotic form of the log of the pdf for some symmetric and Box-Cox symmetric distributions.}
\vspace{-0.8cm}
\begin{center}
\begin{scriptsize}
\begin{tabular}{l|l|l|l|l}
\hline
distribution           & symmetric                     &BCS (${\color{black}{\lambda}}>0$)                         &BCS (${\color{black}{\lambda}}=0$)                        &BCS (${\color{black}{\lambda}}<0$) \\
\hline
normal                 & $k_3=2,$ $k_4=1/(2\sigma^2)$                    &$k_3=2{\color{black}{\lambda}}$, $k_4=1/(2\mu^{2{\color{black}{\lambda}}}\sigma^2{\color{black}{\lambda}}^2)$                                   & $k_1=2$, $k_2=1/(2\sigma^2)$     & $k_1=1$, $k_2=|{\color{black}{\lambda}}|+1$   \\
double exponential     & $k_3=1$, $k_4=\sqrt{2}/\sigma$                  & $k_3={\color{black}{\lambda}}$, $k_4=\sqrt{2}/(\mu^{{\color{black}{\lambda}}}\sigma{\color{black}{\lambda}})$                                  & $k_1=1$, $k_2=\sqrt{2}/\sigma+1$ & $k_1=1$, $k_2=|{\color{black}{\lambda}}|+1$ \\
power exponential      &                                                 &                                                                                 &                                  &                   \\
\hspace{1.5cm} $\tau>1$  & $k_3=\tau,$ $k_4=1/(2 p(\tau)^\tau\sigma^\tau)$ &$k_3={\color{black}{\lambda}}\tau,$ $k_4=1/(2p(\tau)^\tau\mu^{{\color{black}{\lambda}}\tau}\sigma^\tau {\color{black}{\lambda}}^\tau)$          & $k_1=\tau$, $k_2=1/(2p(\tau)^\tau \sigma^\tau)$  & $k_1=1$, $k_2=|{\color{black}{\lambda}}|+1$   \\
\hspace{1.5cm} $\tau=1$  & $k_3=1$, $k_4=\sqrt{2}/\sigma$                  & $k_3={\color{black}{\lambda}}$, $k_4=\sqrt{2}/(\mu^{{\color{black}{\lambda}}}\sigma{\color{black}{\lambda}})$&$k_1=1$, $k_2=\sqrt{2}/\sigma+1$ & $k_1=1$, $k_2=|{\color{black}{\lambda}}|+1$ \\
\hspace{1.5cm} $\tau<1$  & $k_3=\tau,$ $k_4=1/(2 p(\tau)^\tau\sigma^\tau)$ & $k_3={\color{black}{\lambda}}\tau,$ $k_4=1/(2p(\tau)^\tau\mu^{{\color{black}{\lambda}}\tau}\sigma^\tau {\color{black}{\lambda}}^\tau)$         & $k_1=1$, $k_2=1$                 & $k_1=1$, $k_2=|{\color{black}{\lambda}}|+1$   \\
Cauchy                           & $k_1=1,$ $k_2=2$                                & $k_1=1$, $k_2={\color{black}{\lambda}}+1$                                                            & $k_1=1$, $k_2=1$                 & $k_1=1$, $k_2=|{\color{black}{\lambda}}|+1$   \\
t                                    & $k_1=1$, $k_2=\tau+1$                           & $k_1=1$, $k_2={\color{black}{\lambda}}\tau+1$                                                        & $k_1=1$, $k_2=1$                 & $k_1=1$, $k_2=|{\color{black}{\lambda}}|+1$   \\
type I logistic        & $k_3=2$, $k_4=1/\sigma^2$                       & $k_3=2 {\color{black}{\lambda}}$, $k_4=1/(\mu^{2{\color{black}{\lambda}}} \sigma^2 {\color{black}{\lambda}}^2)$                                & $k_1=2$, $k_2=1/\sigma^2$        & $k_1=1$, $k_2=|{\color{black}{\lambda}}|+1$        \\
type II logistic       & $k_3=1$, $k_4=1/\sigma $                        & $k_3={\color{black}{\lambda}}$, $k_4=1/(\mu^{\color{black}{\lambda}} \sigma {\color{black}{\lambda}})$                                         & $k_1=1$, $k_2=1/\sigma+1$        & $k_1=1$, $k_2=|{\color{black}{\lambda}}|+1$ \\
canonical slash          & $k_1=1,$ $k_2=2$                                & $k_1=1$, $k_2={\color{black}{\lambda}}+1$                                                            & $k_1=1$, $k_2=1$                 & $k_1=1$, $k_2=|{\color{black}{\lambda}}|+1$   \\
slash                        & $k_1=1$, $k_2=q+1$                              & $k_1=1$, $k_2={\color{black}{\lambda}} q+1$                                                          & $k_1=1$, $k_2=1$       & $k_1=1$, $k_2=|{\color{black}{\lambda}}|+1$   \\
\hline
\end{tabular}
\end{scriptsize}
\end{center}
\label{tail}
\vspace{-0.3cm}
\end{table}

The study presented in this section shows that the class of the Box-Cox symmetric distributions is very flexible
for modeling positive data displaying different right tail behaviors. It covers from right light-tailed distributions to
heavier than any `Paretian type' tailed distribution. More importantly, some distributions in this class
have an extra parameter that controls the right tail heaviness. Slightly heavy-tailed data may be modeled using a
Box-Cox power exponential distribution, which has a `slightly' heavy right tail (heavy tail but lighter than any `Paretian type' tail)
when ${\color{black}{\lambda}}>0$ and ${\color{black}{\lambda}}\tau<1$ with the tail weight determined by $\tau$. Moderately or highly heavy-tailed data may require a Box-Cox t or Box-Cox slash distribution,
because both are right `Paretian type' tailed distributions with a parameter that controls the right tail heaviness.

{\color{black}{
\section{Likelihood-based inference}
\label{inference}

The log-likelihood function for a single observation $y$ taken from a BCS distribution is given by
$$
\ell(\mu,\sigma,\lambda)= (\lambda-1) \log y -  \lambda \log \mu - \log \sigma 
+\log r(z^2)- \log R\left(\frac{1}{\sigma |\lambda|}\right), \nonumber 
$$
where $z={\color{black}{h}}(y;\mu,\sigma,{\color{black}{\lambda}})$ with ${\color{black}{h(y;\mu,\sigma,\lambda)}}$ given in (\ref{ztrans}); the last term in $\ell$ is zero if $\lambda=0$. The score vector and the Hessian matrix are obtained from the first and second derivatives of $\ell$ with respect to the parameters; see the Appendix. 
}}

{\color{black}{
The maximum likelihood estimates of $\mu$ and $\sigma$, for fixed $\lambda$, from a sample of $n$ independent observations $y_1,\ldots,y_n$, are solution of the system of equations
\begin{equation*}
\mu= \left\{
\begin{array}{ll}
{ \frac{1}{(n\sigma {\color{black}{\lambda}})^{1/{\color{black}{\lambda}}}} \left(\sum_{i=1}^{n} \varpi_i y_{i}^{\color{black}{\lambda}}  z_i\right)^{1/{\color{black}{\lambda}}}}, {\quad \mbox{if} \quad {\color{black}{\lambda}} \neq 0}, \\
{\left(\prod_{i=1}^{n} y_i^{\varpi_i}\right)^{1/\sum_{i=1}^{n} \varpi_i}}, {\quad \mbox{if} \quad {\color{black}{\lambda}}=0},
\end{array}
\right.
\end{equation*}
\begin{equation*}\label{EMV}
\sigma^2=\left\{
\begin{array}{ll}
{ \frac{1}{n{\color{black}{\lambda}}^2(1-\delta)}\sum_{i=1}^{n}\varpi_i\left[\left(\frac{y_i}{\mu}\right)^{{\color{black}{\lambda}}}-1\right]^2}, {\quad \mbox{if} \quad {\color{black}{\lambda}} \neq 0}, \\
{\frac{1}{n} \sum_{i=1}^{n} \varpi_i \left(\log \frac{y_i}{\mu}\right)^2}, {\quad \mbox{if} \quad {\color{black}{\lambda}}=0},
\end{array}
\right.
\end{equation*}
where
$$\delta=\frac{1}{\sigma|{\color{black}{\lambda}}|} \frac{r \left(\frac{1}{\sigma|{\color{black}{\lambda}}|}\right)}{R\left(\frac{1}{\sigma|{\color{black}{\lambda}}|}\right)},$$
$z_i=h(y_i;\mu,\sigma,\lambda)$, and
}}
 $\varpi_i=\varpi(z_i)$ with $\varpi(z)= -2 r'(z^2)/r(z^2)$ being a weighting function that depends on $r$.  
We note that the maximum likelihood estimates of $\mu$ and $\sigma$ involve weighted arithmetic and geometric averages of the contributions of each observation $y_i$ with weight $\varpi(z_i)$. Table (\ref{funcaopeso}) gives $\varpi(z)$ for several BCS distributions.
Note that some distributions in the BCS class produce robust estimation against outliers. For instance,
for the Box-Cox t and Box-Cox power exponential (with $\tau<2$) distributions the weighting function is decreasing in the observation of $Y$. Hence, outlier observations have smaller weights in the estimation of the parameters.

\begin{table}[!ht]
\normalsize \caption{Weighting functions for some BCS distributions.}
\vspace{-0.8cm}
\begin{center}
\begin{tabular}{l|l}
\hline
BCS distribution   & $\varpi(z)$ \\
\hline
normal &  1 \\
double exponential & $\sqrt{2}/|z|$\\
power exponential & $\tau (z^2)^{\tau/2-1}/(2 p(\tau)^{\tau})$\\
Cauchy &   $2/(1+z^2)$   \\
t & $(\tau+1)/(\tau+z^2)$ \\
type I logistic &  $-2(\exp\{-z^2\}-1)/(\exp\{-z^2\}+1)$ \\
type II logistic & $(\exp\{-|z|\}-1)/[|z|(\exp\{-|z|\}+1)]$\\
canonical slash & $2/z^2-\exp\{-z^2/2\}/(1-\exp\{-z^2/2\})$\\
slash & 
          $ 2 \Psi((q+3)/2,z^2/2)/(z^2 \Psi((q+1)/2,z^2/2)) $ \\
\hline
\end{tabular}
\end{center}
\label{funcaopeso}
\end{table}

The system of likelihood equations for $(\mu, \sigma, {\color{black}{\lambda}})$ does not have analytical solution. Furthermore, it may involve an addition equation relative to an
extra parameter (for instance, the degrees of freedom parameter of the BCT distribution). Maximization of the likelihood function is implemented in the package {\tt gamlss} in {\tt R} for the BCT, BCCG and BCPE distributions through the CG and the RS algorithms \citep{Rigby, Rigby4}. It is noteworthy that {\tt gamlss} allows the fit of regression models
with monotonic link functions relating the parameters ($\mu$, $\sigma$, ${\color{black}{\lambda}}$ and the possibly extra parameter) to explanatory variables through parametric or semi-parametric additive models.

{\color{black}{
It is of particular interest to test the null hypothesis ${\rm H}_0: \lambda=0$; recall that the BCS distributions reduce to the log-symmetric distributions when $\lambda=0$. In order to evaluate the performance of the likelihood ratio test of ${\rm H}_0$ against a two sided alternative, we now present a small Monte Carlo simulation study. We set $\mu=\sigma=1$ and generate 10,000 samples of sizes $n=100, 200, 300, 500$ from a Box-Cox t distribution with two different values for the degrees of freedom parameter, namely $\tau=4, 10$. The samples are generated under the null hypothesis. We assume that $\tau$ is known, and we estimate the remaining parameters using the function {\tt optim} in {\tt R} with the analytical derivatives derived in the Appendix and with numerical derivatives. The type I error probability estimated from the simulated samples for a nominal level $\alpha=5\%$ are presented in Table \ref{simultestLR}. The figures in the table reveal that the likelihood ratio test performs well regardless of whether analytical or numerical derivatives are employed. As expected, the type I error probabilities are close to the nominal level of the test for the sample sizes considered here.
}}

\begin{table}[!htp]
\centering
\normalsize 
\caption{{\color{black}{Type I error probability of the likelihood ratio test of ${\rm H}_0: \lambda=0$ using analytical and numerical derivatives; Box-Cox t distribution with $\mu=\sigma=1$ and $\tau=4$ and 10.}}} \label{simultestLR}
\vspace{-0.4cm}
{\color{black}{
\begin{tabular}{c|cc|cc}  \hline
$n$&\multicolumn{2}{c}{$\tau=4$}&\multicolumn{2}{c}{$\tau=10$}\\
\cline{2-5}
   &anal. der.&num. der.&anal. der.&num. der.	        \\\hline
100&	0.043	  &  0.043  &	   0.044 &   0.044           \\ 
200&  0.051   &	 0.051  &    0.052 &   0.052           \\ 
300&	0.050   &	 0.050  &    0.049 &   0.049           \\
500&  0.049   &  0.049  &    0.049 &   0.049           \\
\hline
\end{tabular}
}}
\end{table}

\section{Applications and comparison of alternative approaches}\label{applic}

In this section, we present applications of the Box-Cox distributions in the analysis of micro and macronutrients intake.
The data refer to observations of nutrients intake based on the first 24-hour dietary recall interview for
$n=368$ individuals. 
For each nutrient, we assume that the data $Y_1,\ldots, Y_{n}$ are independent. 

First, we fitted different models to each of all the 33 nutrients. All the models considered in the first analysis are constructed from 
the Student-t distribution and from its limiting case when the degrees of 
freedom parameter goes to infinity, i.e. the normal distribution. The following models were considered:
Box-Cox t (BCT); Box-Cox Cole-Green (BCCG), which corresponds to the BCT model with $\tau\rightarrow \infty$; skew-normal (SN) and skew-t (ST) \citep{azzalini}; 
and transformed symmetric models with normal (TN) and t (TT) errors \citep{gauss}. The TN (TT) model assumes that the original \citet{Box} 
transformed data follow a normal (Student-t) distribution. 
The unknown parameters (including the degrees of freedom parameter) were estimated by the maximum likelihood method.
For the BCT, BCCG, SN and ST distributions, we used the {\tt gamlss} package implemented in {\tt R}, while for the TN and TT models we 
used both the function {\tt optim} in {\tt R} and the {\tt PROC NLP} in {\tt SAS}.
Goodness-of-fit was evaluated using the following criteria: Akaike information criterion (AIC) and Anderson-Darling statistics (AD, ADR, and AD2R); 
see \citet[Tables 1, 2 and B.1]{luceno}. AD is a global measure of lack-of-fit, while both ADR and AD2R are more sensitive to the lack of fit in 
the right tail of the distribution; AD2R puts more weight in the right tail than ADR. For all the four criteria a lower value is preferred.

Tables \ref{aic12}-\ref{ad13} present the goodness-of-fit statistics for all the fitted models to 22 and 11 micro and
macronutrients intakes data. Underlined numbers indicate the best fitting model. The blank cells in the tables indicate that the  algorithm employed for maximum likelihood 
estimation did not achieve convergence or produced unrealistic estimates.
The tables convey important information.
First, the datasets cover a wide range of light-tailed to heavy-tailed data. This can be seen by the estimated values
of the degrees of freedom parameter under the BCT model; $\widehat \tau_{BCT}$ ranges from $2.2$ to $196.0$ (see Table \ref{aic12}).
Second, no convergence problem was observed while fitting the BCT model, the TT model, and the TN model. 
The maximum likelihood estimation under the SN model and under the BCCG model did not achieve convergence 
in $10$ cases. Under the ST model, convergence was not reached in two cases. 
%
%
Third, according to the AIC criterion, the BCT model achieved the best fit in most of the cases (26 out of 33 cases); the AICs of the 
BCT and TT fits coincided in 20 cases. The models derived from the normal distribution, namely the BCCG, SN, and
TN models, did not perform well in general, except for two cases in which the estimated degrees of freedom parameter under the 
BCT model was very large ($\widehat\tau_{BCT}>50$). In such cases, the BCCG and the TN models achieved the best fits.
Forth, according to all the Anderson-Darling criteria, the BCCG, skew-normal and TN models did not outperform the 
BCT, ST and TT models in any of the cases. It suggests that the tail behavior of the nutrients data are better modeled by the 
distributions derived from the t distribution. The BCT and TT models were the best fitting models in most of the cases.
Overall, we conclude that the Box-Cox t model performed better than the other models. The transformed t model behaved as well as the Box-Cox t model in many cases. 
However, as pointed out earlier, the transformed t model has some drawbacks that are overcome by the Box-Cox t model.

\begin{figure}[H]
\centering
\includegraphics[width=0.45\hsize]{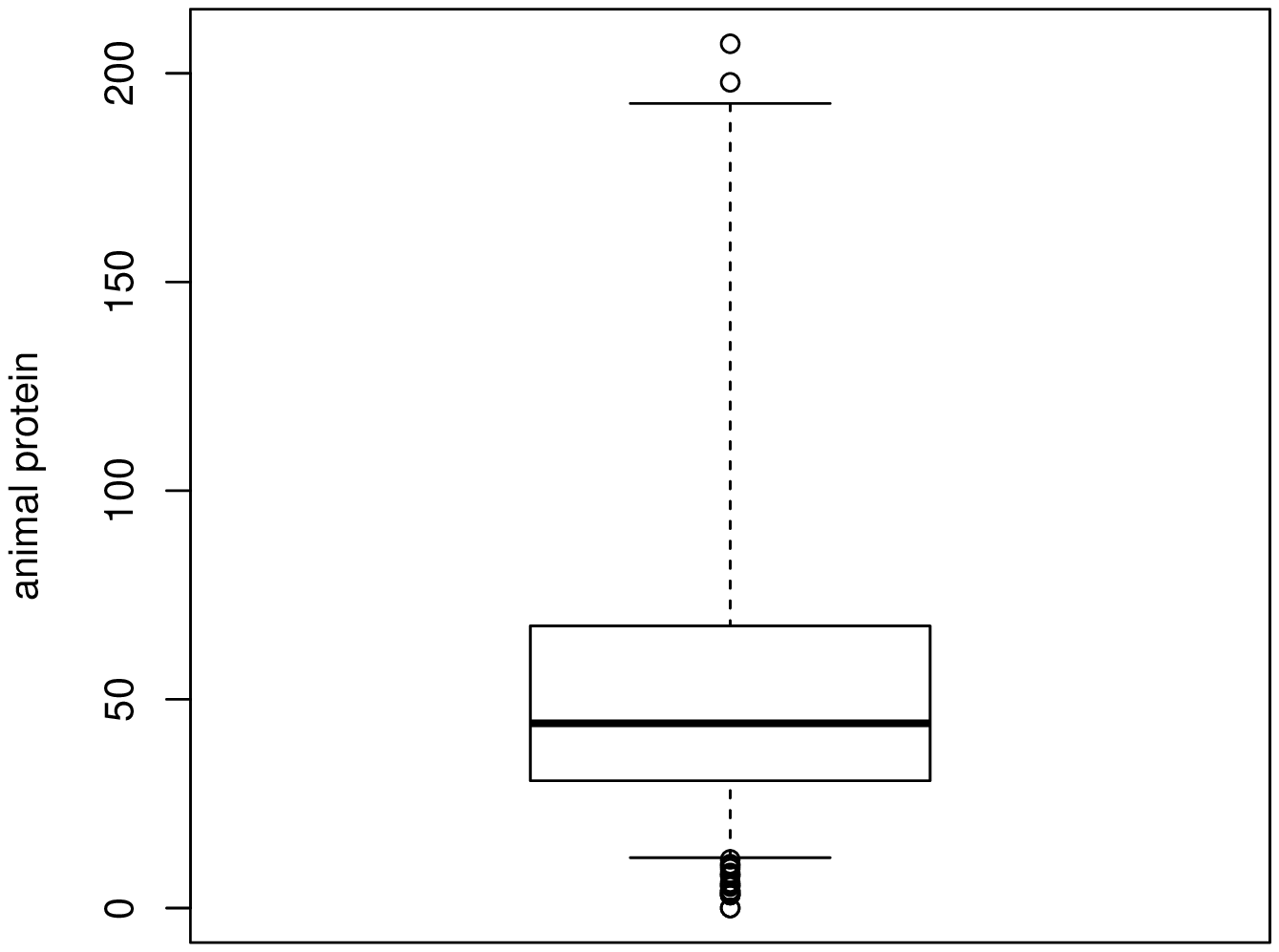}
\includegraphics[width=0.45\hsize]{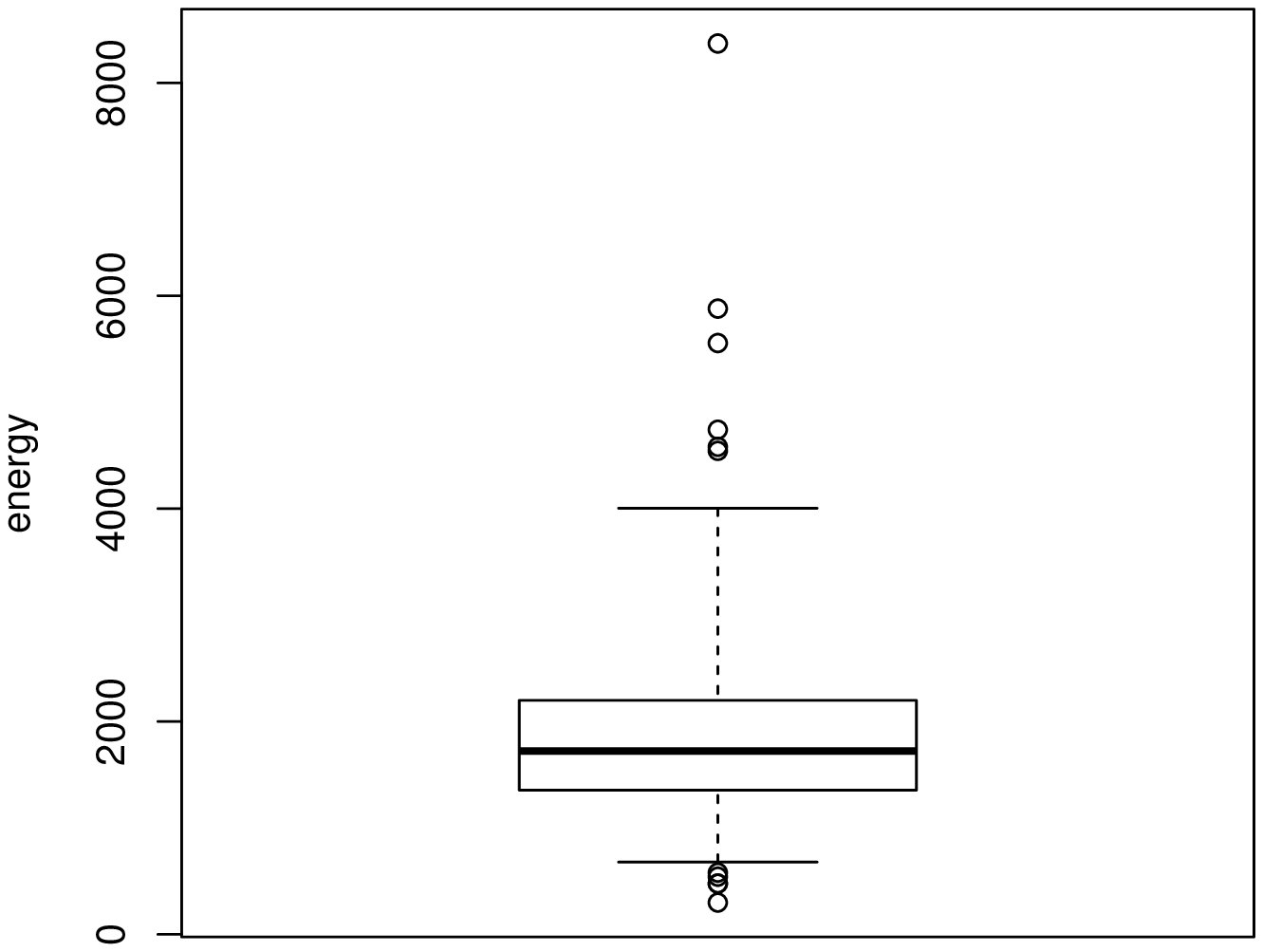}
\caption{Adjusted boxplots; animal protein data (left) and energy intake (right).}
\label{boxplotajustado1}
\end{figure}

We now turn to a detailed analysis of the data on the intake of animal protein and energy. Adjusted boxplots \citep{Hubert} are presented in Figure \ref{boxplotajustado1} 
and show that the data sets are asymmetric and contain outlying observations. It is noteworthy that the data set on energy intake contains an outlier that is well 
above the second highest observed intake. We fitted the Box-Cox t, Box-Cox Cole-Green, Box-Cox power exponential and Box-Cox slash distributions to each data set.
For the BCT, BCCG, and BCPE models, we used the {\tt gamlss} package implemented in {\tt R}, while for the BCSlash model we 
used the function {\tt optim} in {\tt R}.
Tables \ref{protanimal1} and \ref{energia1} give descriptive statistics of the data, and parameter estimates and goodness-of-fit statistics for each fitted model.
The descriptive statistics confirm the findings in the boxplots. For both data sets, the Akaike information criteria are similar for the BCT, BCPE, and BCSlash models, 
and both are smaller than that for the BCCG model. The Anderson-Darling statistics indicate that the BCT model gives the best fit. Note that the BCCG model gives a 
poor right tail fit for both data sets. This lack of fit is most pronounced for the energy intake data set, for which ${\rm AD2R}=112.25$ for the BCCG model while 
${\rm AD2R}=1.78$, ${\rm AD2R}=4.18$, and ${\rm AD2R}=2.15$ for the BCT, BCPE, and BCSlash models, respectively.

\begin{figure}[H]
\centering
\begin{subfigure}[b]{0.24\textwidth}
    \includegraphics[width=1 \hsize]{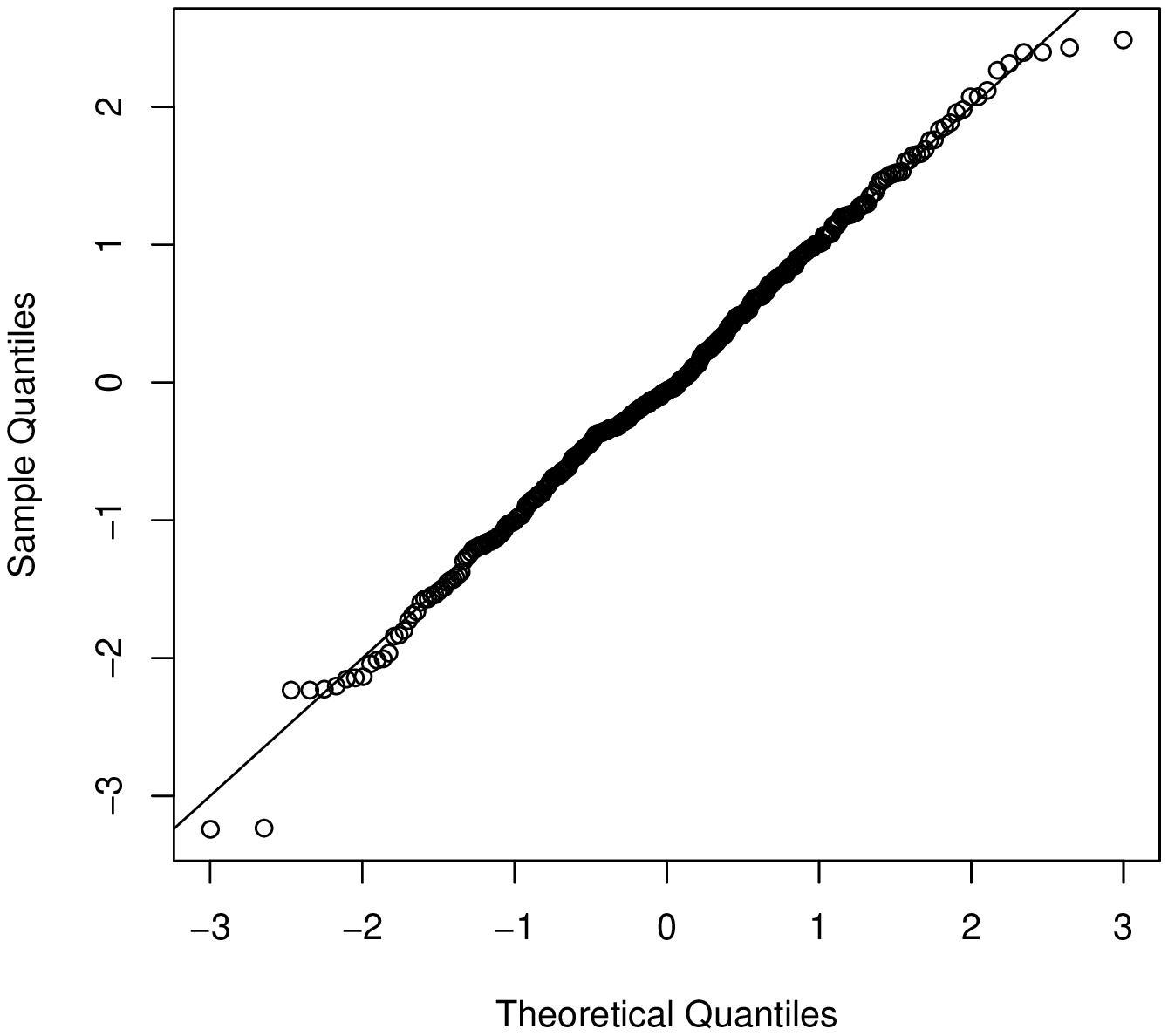}
                \caption{BCT}
                \label{residuobctprotanimaliid}
        \end{subfigure}
\begin{subfigure}[b]{0.24\textwidth}
    \includegraphics[width=1 \hsize]{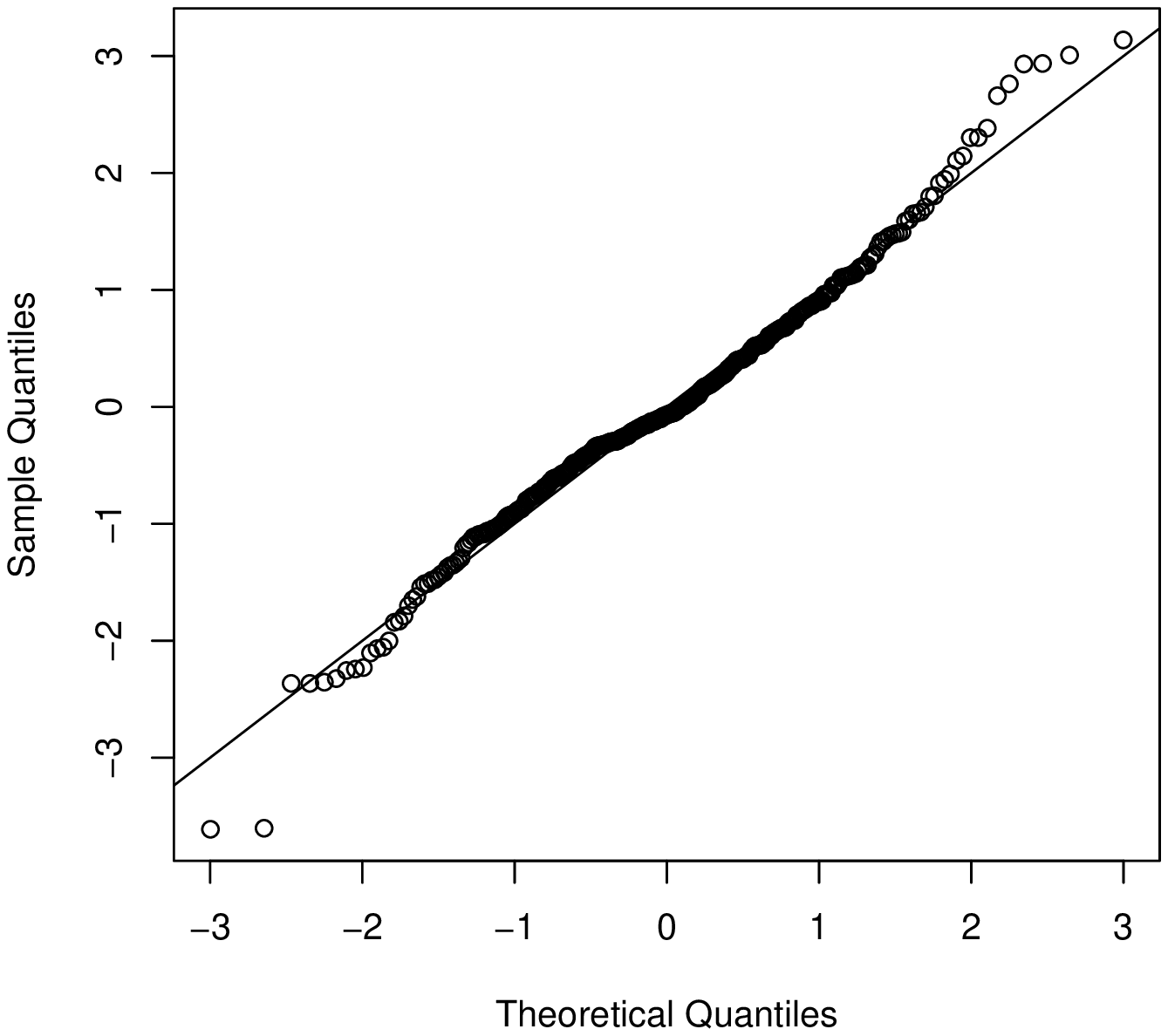}
                \caption{BCCG}
                \label{residuobccgprotanimaliid}
        \end{subfigure}
 \begin{subfigure}[b]{0.24\textwidth}
    \includegraphics[width=1 \hsize]{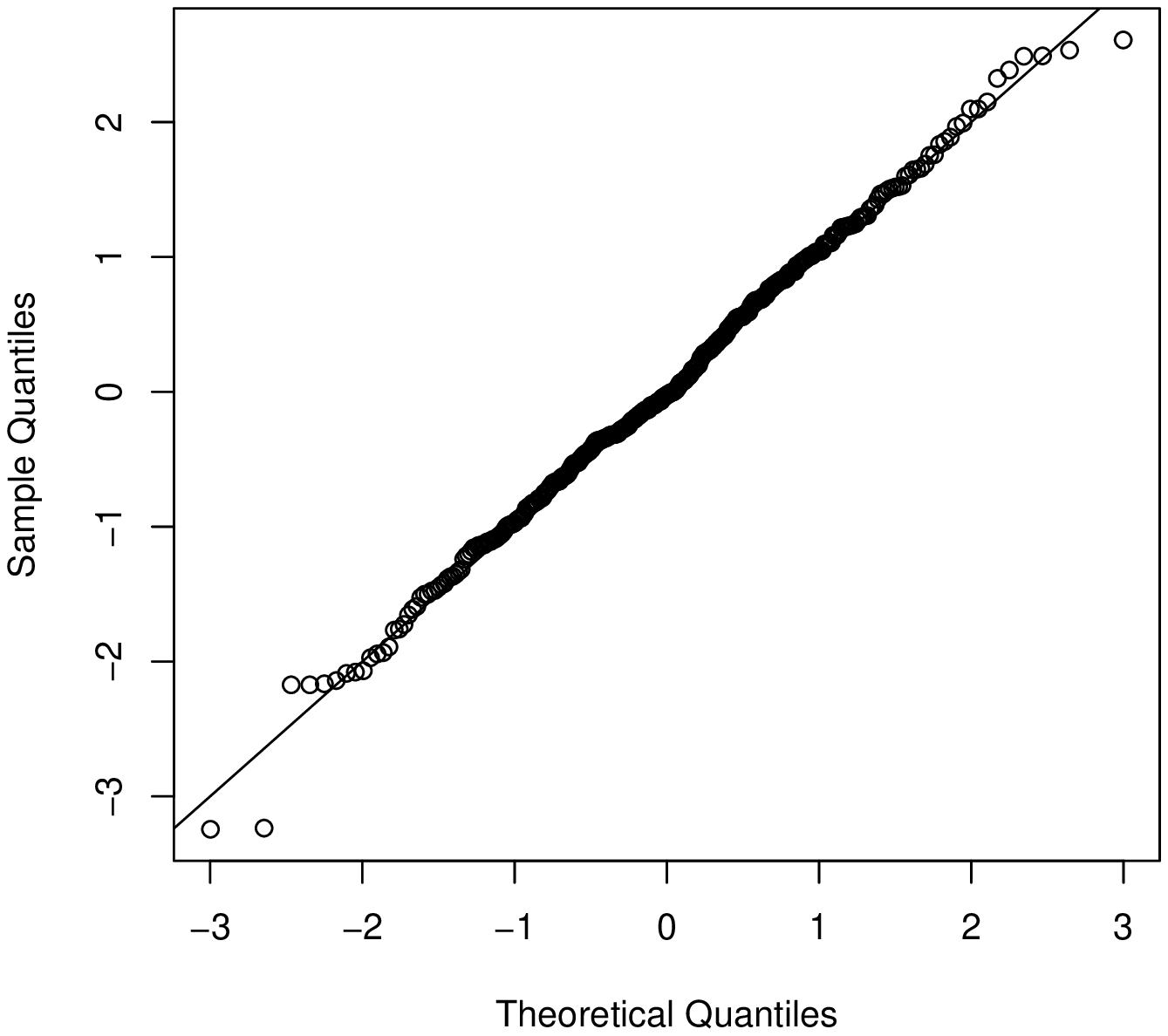}
                \caption{BCPE}
                \label{residuobcpeprotanimaliid}
        \end{subfigure}
 \begin{subfigure}[b]{0.24\textwidth}
    \includegraphics[width=1 \hsize]{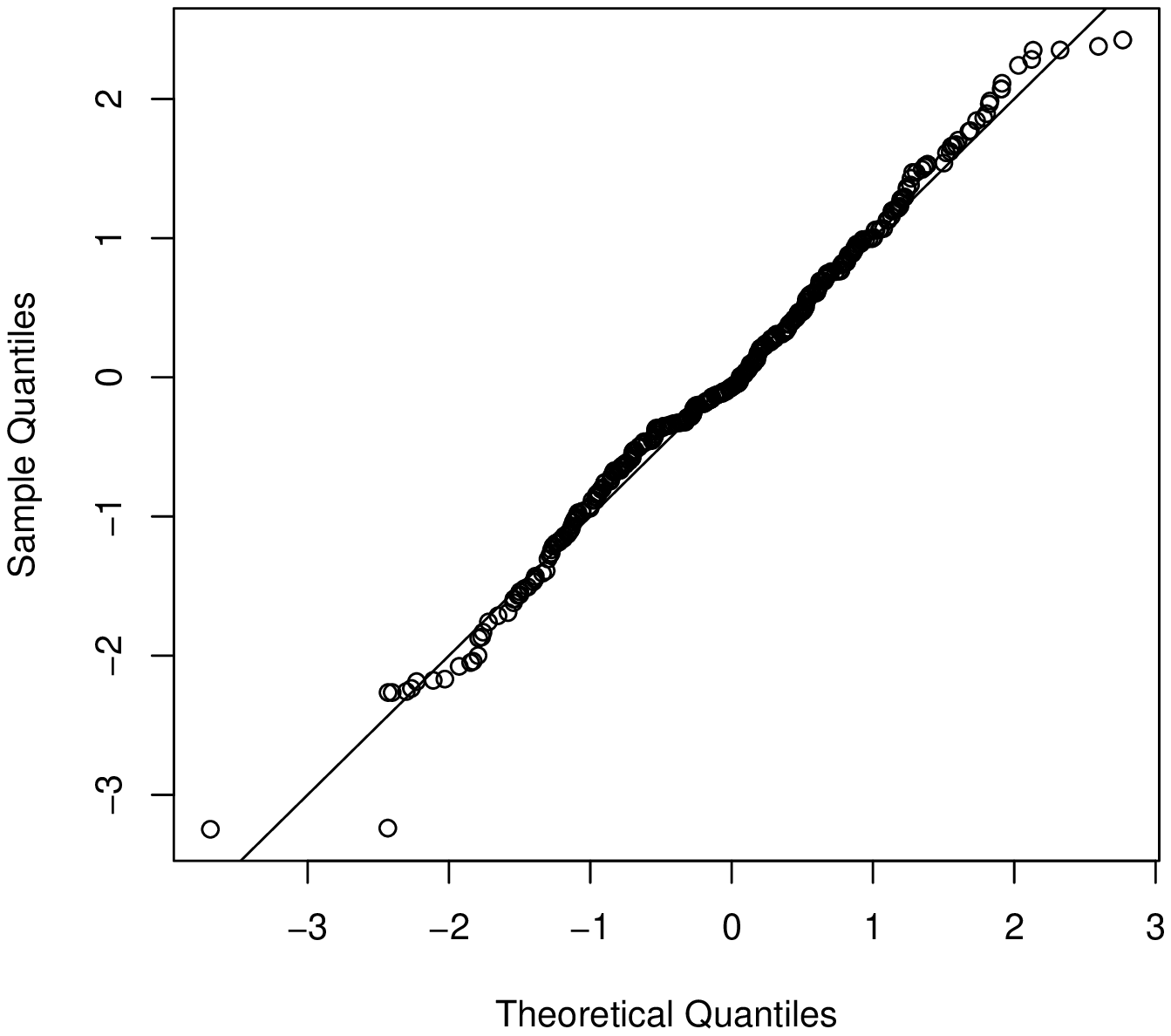}
                \caption{BCSlash}
                \label{residuobcslashprotanimaliid}
        \end{subfigure}
\begin{subfigure}[b]{0.24\textwidth}
    \includegraphics[width=1 \hsize]{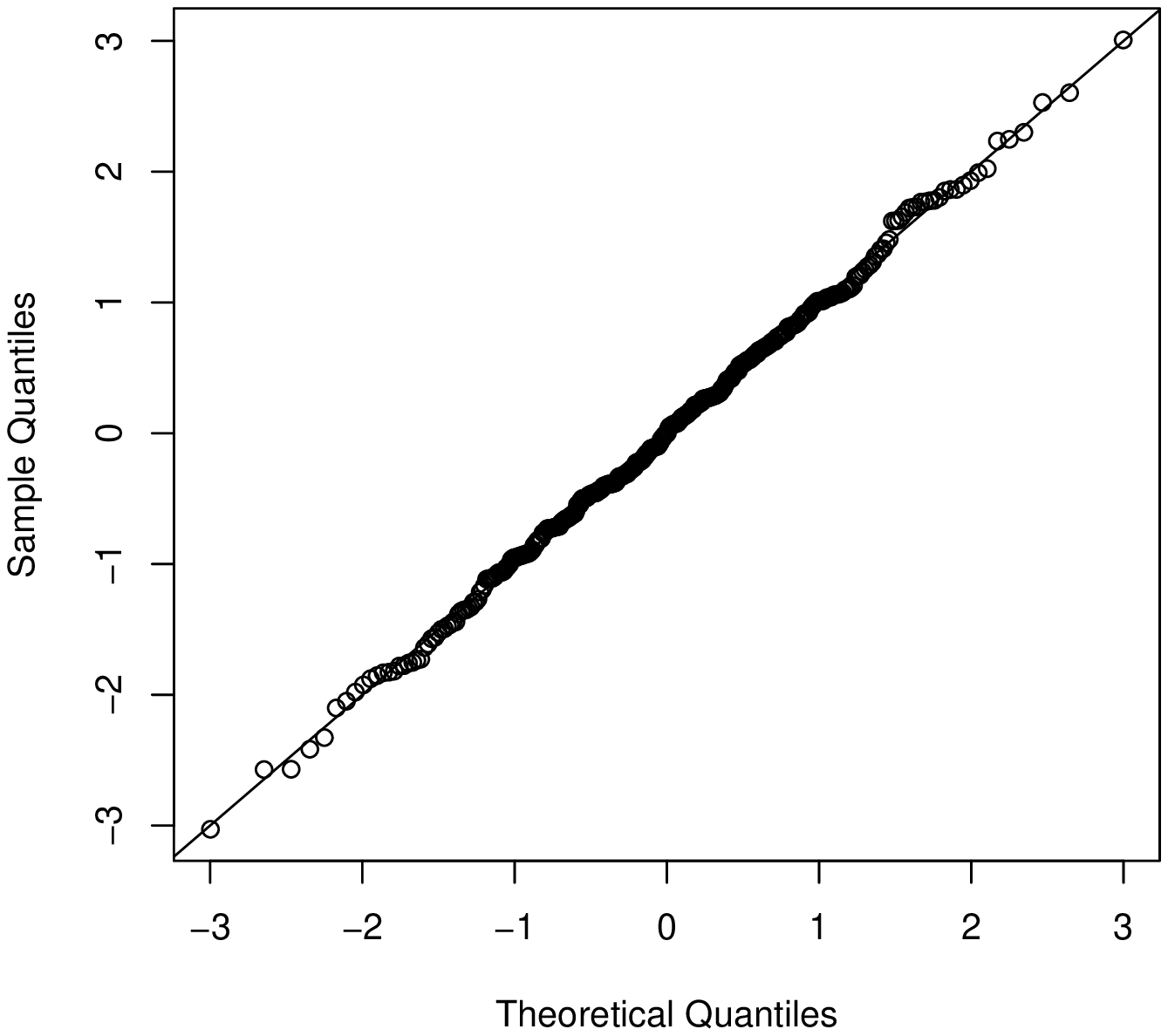}
                \caption{BCT}
                \label{residuobctenergiaiid}
        \end{subfigure}
\begin{subfigure}[b]{0.24\textwidth}
    \includegraphics[width=1 \hsize]{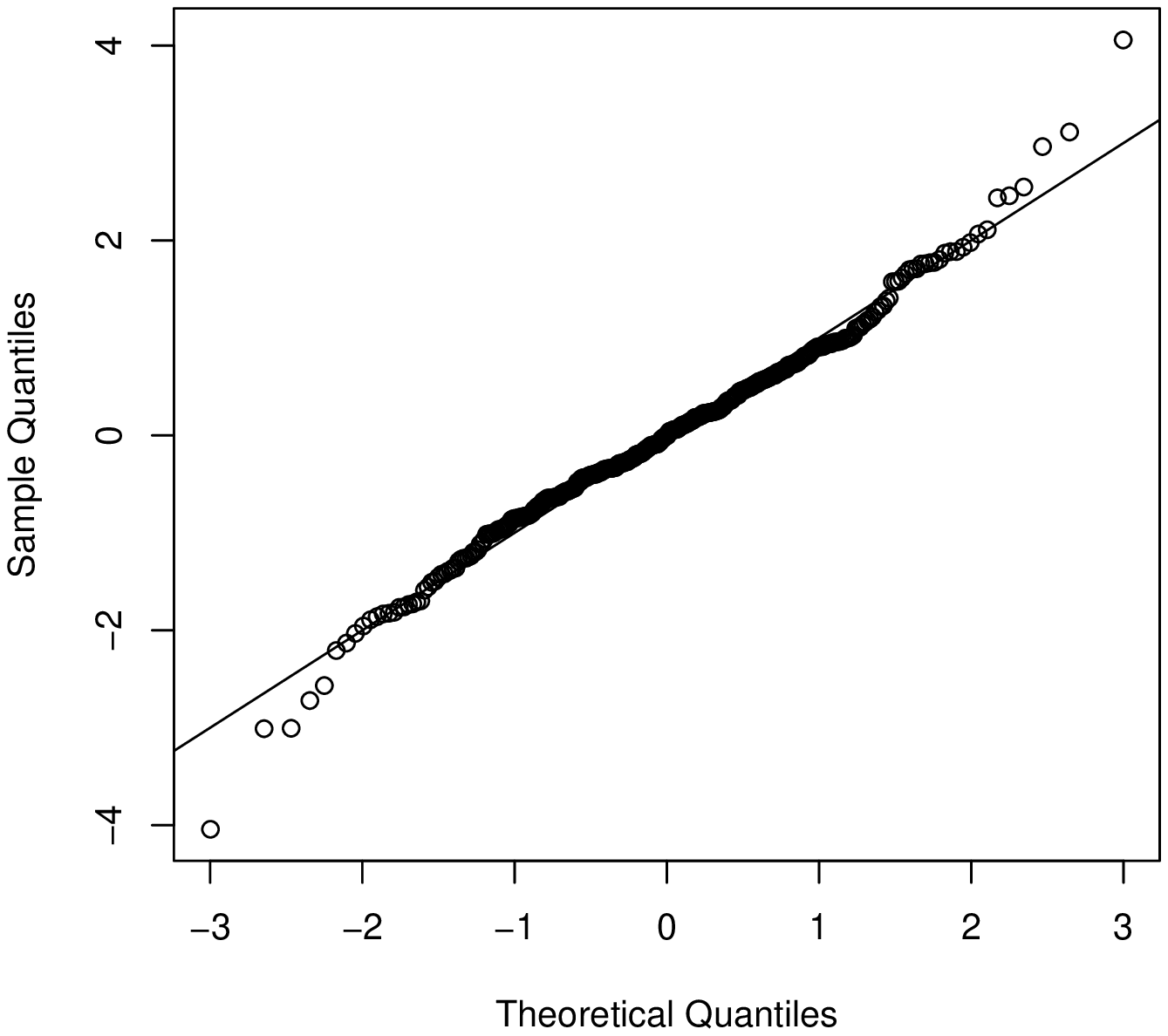}
                \caption{BCCG}
                \label{residuobccgenergiaiid}
        \end{subfigure}
 \begin{subfigure}[b]{0.24\textwidth}
    \includegraphics[width=1 \hsize]{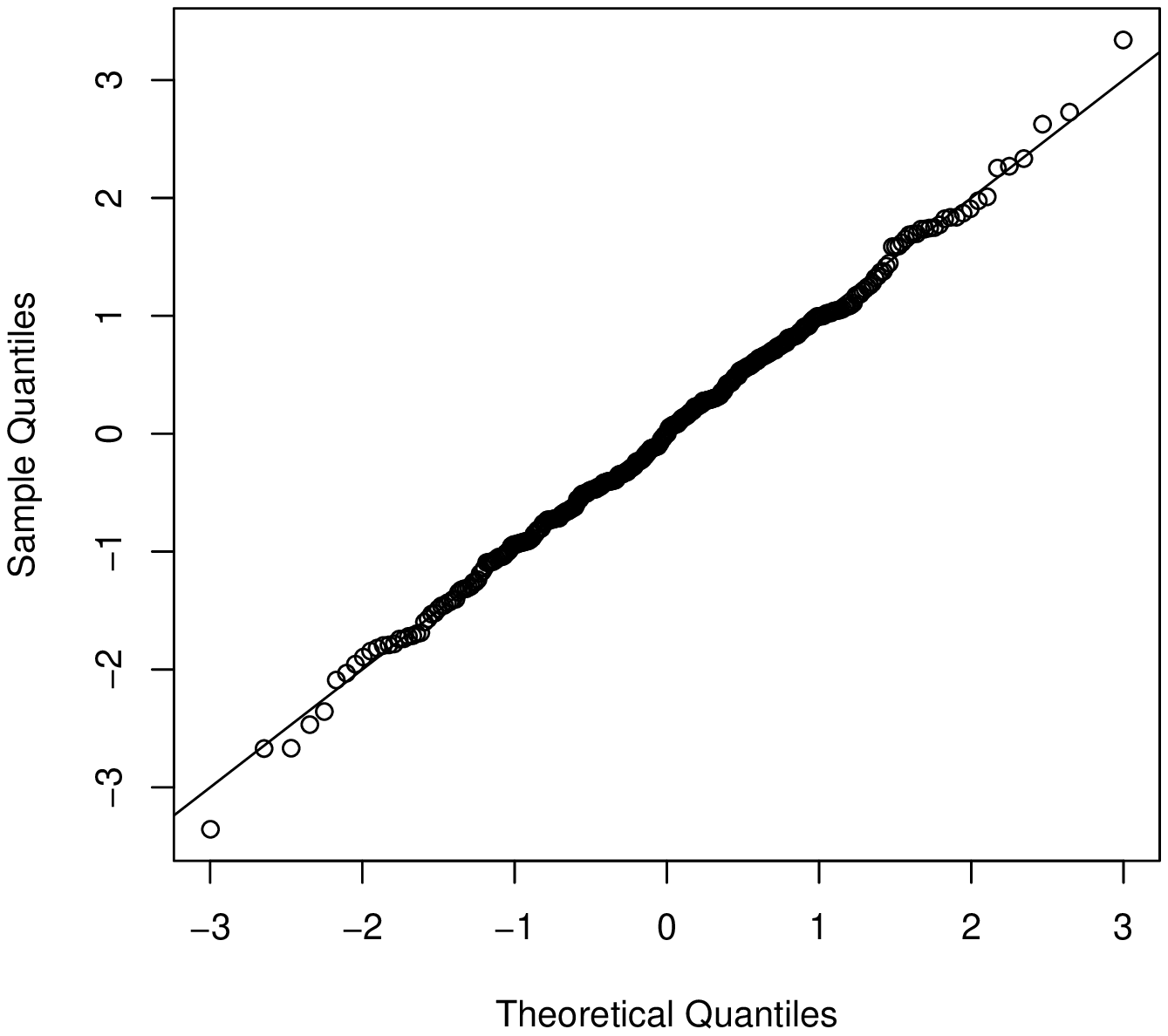}
                \caption{BCPE}
                \label{residuobcpeenergiaiid}
        \end{subfigure}
         \begin{subfigure}[b]{0.24\textwidth}
\includegraphics[width=1 \hsize]{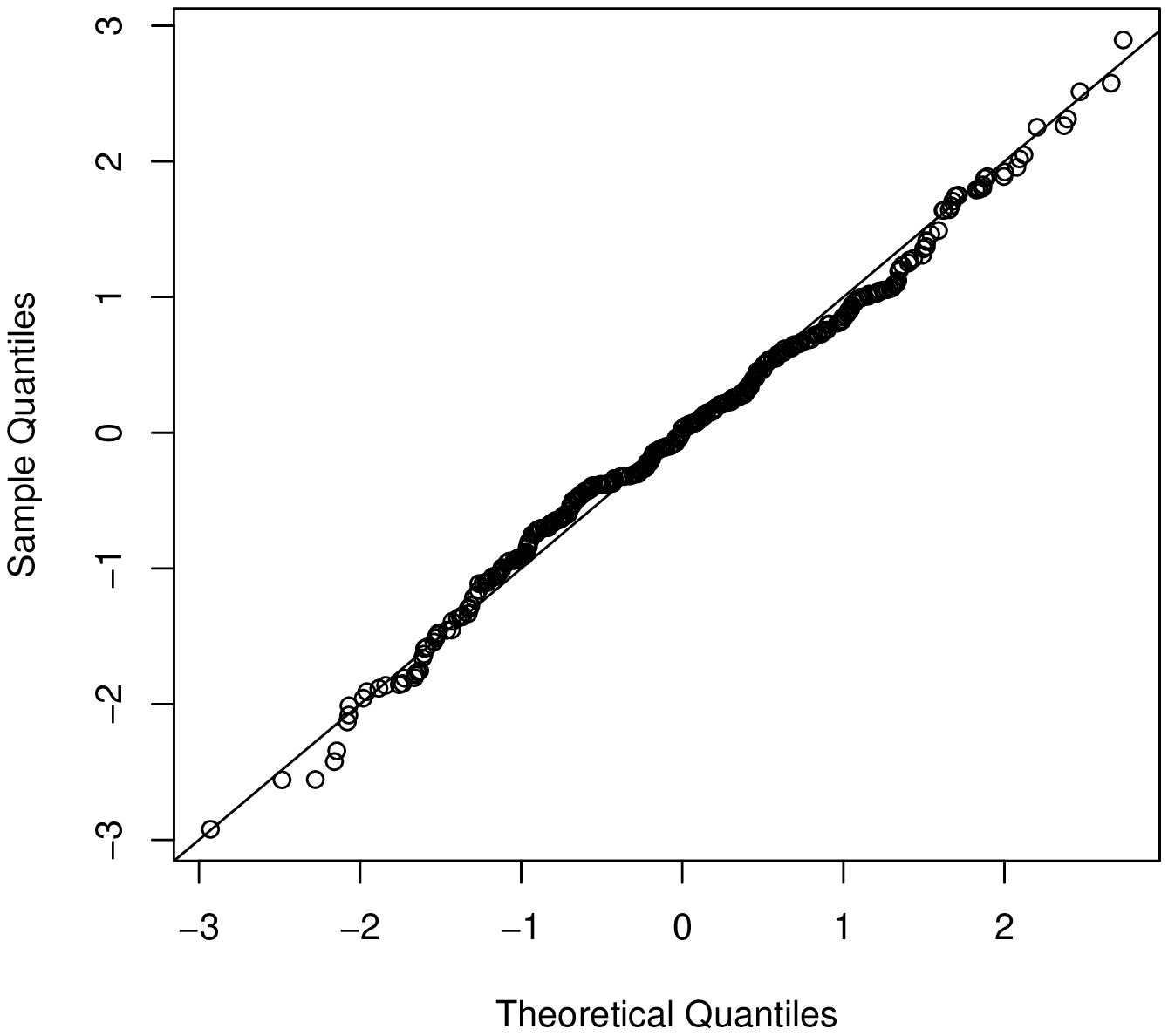}
                \caption{BCSlash}
                \label{residuobcslashenergiaiid}
        \end{subfigure}
\caption{qq plots for quantile residuals for the BCT, BCCG, BCPE and BCSlash model fits for the animal protein intake data (first row) and energy intake data (second row).}
\label{residuos}
\end{figure}

Figure \ref{residuos} presents qq plots for quantile residuals $\widehat r = \Phi^{-1}(\widehat F_Y(y))$, where $\Phi(\cdot)$ is the cdf of the standard normal distribution and
$\widehat F_Y(y)$ is the fitted cdf of $Y$. If the model is correct, the quantile residuals are expected to behave approximately as standard normal quantiles \citep{Dunn}.
The lack of fit of the BCCG model in the right tail is
clear from the plots for both data sets. For the animal protein data, the residual plots are similar for the BCT and BCPE models and indicate reasonable fits.
On the other hand, for the energy intake data set, which seems to require a right heavier tailed model, the BCT model provides a slightly better fit than the BCPE and BCSlash models.

\begin{table}[H]
\normalsize \caption{AIC for the fitted BCT, ST, TT, BCCG, SN, and TN models; micronutrients and macronutrients datasets.}
\vspace{-0.8cm}
\footnotesize
\begin{center}
\begin{tabular}{l|r|r|r|r|r|r|r}
\hline
 micronutrient      &$\widehat{\tau}_{BCT}$&        BCT    & ST      & TT             & BCCG    & SN      & TN  \\
\hline
vitamin A (mcg)      & 7.2 &\underl{5807.4}&         5809.2 & \underl{5807.4}&         &         &       5822.7\\
vitamin D (mcg)      & 6.8 &\underl{1688.7}&         1707.7 &         1690.5 &         &  1745.3 &       1698.9\\
vitamin E (mg)       & 6.9 &\underl{1812.8}&         1818.4 & \underl{1812.8}&  1824.3 &  1902.4 &       1824.3\\
vitamin K   (mcg)    & 7.8 &\underl{4354.3}&         4356.5 &\underl{4354.3} &         &         &       4368.4\\
vitamin C   (mg)     & 2.2 &\underl{4022.5}&                &         4029.9 &         &         &       4120.9\\
vitamin B1 (mg)      & 6.8 &\underl{709.7} &          711.6 &  \underl{709.7}&   720.9 &  778.4  &        720.9\\
vitamin B2  (mg)     & 6.5 &\underl{701.6} &          702.2 &  \underl{701.6}&   716.1 &  769.6  &        716.1\\
vitamin B3 (mg)      & 6.8 &\underl{2722.4}&          2726.3&\underl{2722.4} &  2730.8 &  2778.9 &       2730.8\\
vitamin B6  (mg)     &51.3 &         853.2 &           858.5&          853.2 &\underl{851.3} & 866.9&\underl{851.3}\\
vitamin B12 (mcg)    & 2.5 &\underl{1782.6}&         1808.4 &        1782.8 &         &        &       1887.3\\
pantothenic acid (mg)& 8.3 &        1466.0 &\underl{1465.5} &        1466.0 &  1474.1 &  1509.3 &       1474.1\\
folate  (mcg)        & 4.9 &\underl{4795.4}&         4800.7 &\underl{4795.4}& 4823.4  &         &       4823.4\\
calcium (mg)         & 13.3&\underl{5311.3}&         5312.6 &\underl{5311.3}& 5312.5  &  5342.8 &       5312.5\\
phosphorus (mg)      & 14.7&        5548.5 &\underl{5548.4} &    5548.5& 5549.0  &  5561.5 &       5549.0\\
magnesium (mg)       & 8.6 &\underl{4474.8}&         4476.8 &\underl{4474.8}& 4481.5  &  4522.8 &       4481.5\\
iron    (mg)         & 5.9 &        2409.7 &\underl{2409.3} &         2409.7& 2431.9  &         &       2431.9\\
zinc (mg)            & 14.6&\underl{2185.3}&         2188.7 &\underl{2185.3}& 2185.4  &  2203.4 &       2185.4\\
copper (mg)          & 5.5 &\underl{566.1} &                &\underl{566.1} &         &         &        593.6\\
selenium (mcg)       & 5.2 &\underl{3992.6}&         4000.6 &\underl{3992.6}&         &         &       4020.8\\
sodium (mg)          & 4.6 &\underl{6525.4}&         6535.6 &\underl{6525.4}&         &         &       6572.3\\
potassium (mg)       & 9.3 &\underl{6144.5}&         6147.5 &\underl{6144.5}& 6151.9  &  6195.2 &       6151.9\\
manganese (mg)       & 2.5 &\underl{1616.5}&         1647.0 &        1623.7 &         &         &       1656.2\\
\hline
\hline
 macronutrient      &$\widehat{\tau}_{BCT}$&        BCT    & ST      & TT             & BCCG    & SN      & TN  \\
\hline
protein (g)             & 10.2 &\underl{3659.5}&        3660.9 &\underl{3659.5}&  3662.5      & 3678.8 &         3662.5\\
energy (kcal)           & 6.1  &\underl{5861.9}&        5863.7 &\underl{5861.9}&  5876.1      & 5912.8 &         5876.1\\
fiber (g)               & 10.0 &\underl{2652.2}&        2653.5 &\underl{2652.2}&  2655.6      & 2669.2 &         2655.6\\
carbohydrate (g)        & 10.5 &        4360.0 &\underl{4359.9}&         4360.0&  4366.5      & 4382.3 &         4366.5\\
total fat (g)           & 13.9 &\underl{3587.0}&        3593.7 &\underl{3587.0}&  3587.5      & 3651.9 &         3587.5\\
animal protein (g)      & 4.9  &\underl{3514.4}&        3532.5 &         3516.1&  3526.3      & 3553.9 &         3526.5\\
vegetable protein (g)   & 6.6  &\underl{2963.3}&        2964.4 &\underl{2963.3}&  2972.6      & 2988.5 &         2972.6\\
saturated fat (g)       & 196.0&         2819.6&        2822.9 &         2819.6&\underl{2817.7}& 2844.2 &\underl{2817.7}\\
monounsaturated fat (g) & 12.6 &\underl{2857.1}&        2865.4 &\underl{2857.1}&  2858.4      & 2920.1 &         2858.4\\
polyunsaturated fat (g) & 7.0  &         2596.9&\underl{2596.4}&         2596.9&  2612.4      & 2771.9 &         2612.4\\
cholesterol (mg)        & 5.8  &\underl{4724.7}&        4753.7 &         4725.6&              & 4782.9 &         4728.5\\
\hline
\end{tabular}
\end{center}
\label{aic12}
\end{table}

\begin{table}[H]
\normalsize \caption{Anderson-Darling statistics for the fitted BCT, ST, TT, BCCG, SN, and TN models; micronutrients datasets.}
\vspace{-0.8cm}
\footnotesize
\begin{center}
\begin{tabular}{l|r|c|r|r|r|r|r}
\hline
 micronutrient               &statistic&        BCT    & ST      & TT             & BCCG    & SN      & TN  \\
\hline
\multirow{3}{*}{vitamin A}        &AD  &          1.12 & \underl{0.94} &          1.12 &        &       &  1.47 \\
                                  &ADR &          0.64 & \underl{0.53} &          0.64 &        &       &  0.94 \\
                                  &AD2R& \underl{6.26} &          8.46 &          6.27 &        &       & $>100$ \\
\hline
\multirow{3}{*}{vitamin D}        &AD  & \underl{0.25} &          0.67 &          0.28 &        & 3.30  & 0.64 \\
                                  &ADR & \underl{0.11} &          0.32 &          0.13 &        & 2.27  & 0.38 \\
                                  &AD2R& \underl{3.40} &         13.95 &          3.80 &        & $>100$& $>100$ \\
\hline
\multirow{3}{*}{vitamin E}        &AD  & \underl{0.21} &          0.30 & \underl{0.21} &  0.91  &  7.62 &  0.91 \\
                                  &ADR & \underl{0.13} &          0.18 & \underl{0.13} &  0.52  &  5.17 &  0.52 \\
                                  &AD2R& \underl{3.39} &          8.58 & \underl{3.39} & 61.43  & $>100$& 61.43 \\
 \hline
\multirow{3}{*}{vitamin K}        &AD  & \underl{0.60} &         22.69 & \underl{0.60} &        &       &  1.28 \\
                                  &ADR & \underl{0.35} &         2.22  & \underl{0.35} &        &       &  0.86 \\
                                  &AD2R&         55.60 &        $>100$ & \underl{55.44}&        &       & $>100$ \\
 \hline
 \multirow{3}{*}{vitamin C}       &AD  & \underl{0.70} &               &          0.71 &        &       &   7.92 \\
                                  &ADR & \underl{0.34} &               &          0.39 &        &       &   4.23 \\
                                  &AD2R& \underl{6.28} &               &          6.35 &        &       & $>100$ \\
\hline
 \multirow{3}{*}{vitamin B1}      &AD&           0.20  & \underl{0.17} &          0.20 &   1.02 & 6.29 &   1.02 \\
                                  &ADR& \underl{0.10}  & \underl{0.10} &         0.11  &   0.47 & 3.96 &   0.47 \\
                                  &AD2R& \underl{2.83} &          4.36 & \underl{2.83} & $>100$ & $>100$& $>100$ \\
\hline
 \multirow{3}{*}{vitamin B2}      &AD& \underl{0.24}   &  \underl{0.24}& \underl{0.24} &   1.02 &  5.90 &   1.02 \\
                                  &ADR& \underl{0.13}  &          0.14 & \underl{0.13} &   0.46 &  3.79 &   0.47\\
                                  &AD2R& \underl{2.10} &          4.36 & \underl{2.10} & $>100$ & $>100$& $>100$\\
\hline
\multirow{3}{*}{vitamin B3}       &AD  & \underl{0.22} &          0.30 & \underl{0.22} &  1.07  &  6.54 &  1.07 \\
                                  &ADR & \underl{0.15} &          0.18 & \underl{0.15} &  0.56  &  4.18 &  0.56 \\
                                  &AD2R& \underl{3.46} &          4.23 & \underl{3.46} & 17.29  & $>100$& 17.25 \\
\hline
\multirow{3}{*}{vitamin B6}       &AD  & \underl{0.37} &          0.42 &          0.38 &   0.45 &  1.69 & 0.45 \\
                                  &ADR & \underl{0.18} &          0.21 &  \underl{0.18}&   0.21 &  1.04 & 0.21 \\
                                  &AD2R&         4.13  &          5.00 &  \underl{4.10}&   4.44 &  27.4 & 4.45 \\
 \hline
\multirow{3}{*}{vitamin B12}      &AD  & \underl{0.36} &         1.23  &          0.37 &        &       &   8.69 \\
                                  &ADR & \underl{0.26} &         0.77  &          0.27 &        &       &   5.08 \\
                                  &AD2R& \underl{8.88} &         22.56 &          9.67 &        &       & $>100$ \\
\hline
\multirow{3}{*}{panthotenic acid} &AD & \underl{0.24}  &          0.26 & \underl{0.24} &  0.65  &  4.59 &  0.65 \\
                                  &ADR& \underl{0.10}  &          0.16 & \underl{0.10} &  0.34  &  3.10 &  0.34 \\
                                  &AD2R& \underl{2.38} &          2.98 & \underl{2.38} &  3.77  & $>100$& 13.77 \\
\hline
\multirow{3}{*}{folate}           &AD &  \underl{0.19} &          0.23 & \underl{0.19} &   1.83 &       &   1.83 \\
                                  &ADR&  \underl{0.12} &          0.13 & \underl{0.12} &   0.94 &       &   0.95 \\
                                  &AD2R&          4.61 &         14.68 & \underl{4.60}  & $>100$ &       & $>100$ \\
\hline
\multirow{3}{*}{calcium}          &AD  &          0.28 & \underl{0.23} &          0.28 &   0.48 &  4.06 & 0.48 \\
                                  &ADR & \underl{0.11} & \underl{0.11} & \underl{0.11} &   0.18 &  2.38 & 0.18 \\
                                  &AD2R& \underl{2.43} &          2.52 & \underl{2.43} &   7.49 & $>100$& 7.50 \\
\hline
\end{tabular}
\end{center}
\label{ad11}
\end{table}

 \begin{table}[H]
\normalsize \caption{Anderson-Darling statistics for the fitted BCT, ST, TT, BCCG, SN, and TN models; micronutrients datasets ({\it cont.}).}
\vspace{-0.8cm}
\footnotesize 
\begin{center}
\begin{tabular}{l|r|c|r|r|r|r|r}
\hline
 micronutrient         &statistic&        BCT    & ST            & TT            & BCCG    & SN      & TN  \\
\hline
\multirow{3}{*}{phosphorus} &AD  &          0.25 & \underl{0.21} &          0.25 &  0.36 &  1.50 & 0.36 \\
                            &ADR &          0.15 & \underl{0.13} &          0.15 &  0.18 &  1.00 & 0.19 \\
                            &AD2R& \underl{2.42} &          2.94 &          2.43 &  5.58 &$>100$ & 5.56 \\
\hline
\multirow{3}{*}{magnesium}  &AD  & \underl{0.31} &          0.32 & \underl{0.31} &  0.66 &  4.07 &  0.66 \\
                            &ADR & \underl{0.18} &          0.20 & \underl{0.18} &  0.36 &  2.77 &  0.36 \\
                            &AD2R& \underl{2.76} &          5.08 & \underl{2.76} & 25.34 &$>100$ & 25.36 \\
\hline
\multirow{3}{*}{iron}       &AD  &          0.25 & \underl{0.12} &          0.25 &  1.21 &       &   1.21 \\
                            &ADR &          0.10 & \underl{0.07} &          0.10 &  0.45 &       &   0.45 \\
                            &AD2R&          2.35 &          8.01 &  \underl{2.34}&$>100$ &       & $>100$ \\
 \hline
\multirow{3}{*}{zinc}       &AD  & \underl{0.16} &          0.19 & \underl{0.16} &  0.38 &  1.58 & 0.38 \\
                            &ADR & \underl{0.09} &          0.10 & \underl{0.09} &  0.19 &  1.04 & 0.19 \\
                            &AD2R&         1.92  &          2.96 & \underl{1.91} &  8.28 &$>100$ & 7.97 \\
\hline
\multirow{3}{*}{copper}     &AD  & \underl{0.37} &               & \underl{0.37} &       &       &  1.74 \\
                            &ADR & \underl{0.21} &               & \underl{0.21} &       &       &  0.99 \\
                            &AD2R& \underl{16.10}&               &        16.25  &       &       & $>100$ \\
 \hline
\multirow{3}{*}{selenium}   &AD  & \underl{0.21} &          0.28 & \underl{0.21} &       &       &   1.70 \\
                            &ADR & \underl{0.12} &          0.17 & \underl{0.12} &       &       &   0.83 \\
                            &AD2R&          7.22 &        $>100$ & \underl{7.14} &       &       & $>100$ \\
\hline
\multirow{3}{*}{sodium}     &AD  & \underl{0.18} &          0.41 & \underl{0.18} &       &       &   2.28 \\
                            &ADR & \underl{0.09} &          0.20 & \underl{0.09} &       &       &   1.24 \\
                            &AD2R&         8.28  &         99.13 & \underl{8.27}  &       &       & $>100$ \\
\hline
\multirow{3}{*}{potassium}  &AD  & \underl{0.34} &          0.37 & \underl{0.34} &  0.57 &  2.90 & 0.57 \\
                            &ADR & \underl{0.22} &          0.25 & \underl{0.22} &  0.34 &  2.07 & 0.34 \\
                            &AD2R&        10.71  &         24.71 & \underl{10.68}&$>100$ & $>100$& $>100$ \\
\hline
\multirow{3}{*}{manganese}  &AD  &          1.29 &         2.23  & \underl{0.92} &       &       &  4.64 \\
                            &ADR &          1.03 &         1.67  & \underl{0.69} &       &       &  2.85 \\
                            &AD2R&        $>100$ &        74.60  &\underl{28.29} &       &       & 56.04 \\
\hline
\end{tabular}
\end{center}
\label{ad12}
\end{table}

\begin{table}[H]
\normalsize \caption{Anderson-Darling statistics for the fitted BCT, ST, TT, BCCG, SN, and TN models; macronutrients datasets.}
\vspace{-0.8cm}
\footnotesize 
\begin{center}
\begin{tabular}{l|r|c|r|r|r|r|r|r|r|r}
\hline
 macronutrient                 &statistics&         BCT    & ST            & TT             & BCCG    & SN      & TN  \\
\hline
\multirow{3}{*}{protein}             &AD  &  \underl{0.23} &          0.25 & \underl{0.23} &          0.54 &   2.97 & 0.54 \\
                                     &ADR &  \underl{0.14} &          0.15 & \underl{0.14} &          0.30 &   2.02 & 0.30 \\
                                     &AD2R&          3.60  &          4.62 & \underl{3.58} &         10.93 &  67.43 & 10.93 \\
\hline
\multirow{3}{*}{energy}              &AD&    \underl{0.19} &          0.20 & \underl{0.19} &          1.13 &   4.13 &   1.10 \\
                                     &ADR&   \underl{0.10} &          0.11 & \underl{0.10} &          0.58 &   2.78 &   0.54 \\
                                     &AD2R&  \underl{1.78} &          2.60 & \underl{1.78} &        $>100$ & $>100$ & $>100$  \\
\hline
\multirow{3}{*}{fiber}               &AD&    \underl{0.21} & \underl{0.21} & \underl{0.21} &          0.51 &   1.72 &  0.51 \\
                                     &ADR&   \underl{0.11} &          0.13 & \underl{0.11} &          0.26 &   1.20 &  0.26 \\
                                     &AD2R&          2.03  &          2.50 & \underl{2.02} &         11.28 & $>100$ & 11.26 \\
\hline
\multirow{3}{*}{carbohydrate}        &AD  &           0.26 & \underl{0.20} &          0.25 &          0.39 &  1.36  &   0.39 \\
                                     &ADR &           0.13 & \underl{0.12} &          0.13 &          0.17 &  0.91  &   0.17 \\
                                     &AD2R&           5.22 &         16.34 & \underl{5.21} &        $>100$ & $>100$ & $>100$ \\
\hline
 \multirow{3}{*}{total fat}          &AD  &  \underl{0.41} &          0.52 & \underl{0.41} &          0.62 &   5.66 &  0.62 \\
                                     &ADR &  \underl{0.24} &          0.27 &         0.25  &          0.36 &   3.77 &  0.36 \\
                                     &AD2R&          5.70  &          11.12& \underl{5.54} &         14.46 &  $>100$& 14.46 \\
\hline
  \multirow{3}{*}{animal protein}    &AD&    \underl{0.35} &          0.42 & \underl{0.35} &          1.46 &   2.84 &  1.43 \\
                                     &ADR&            0.16 &          0.16 & \underl{0.14} &          0.70 &   1.74 &  0.68 \\
                                     &AD2R&           3.19 &          4.10 & \underl{3.03} &         23.47 & $>100$ & 22.34 \\
\hline
 \multirow{3}{*}{vegetable protein}  &AD  &          0.25  &  \underl{0.10} &          0.25 &          1.08 &  2.58  & 1.08 \\
                                     &ADR &          0.10  &  \underl{0.07}&          0.10 &          0.43 &  1.62  & 0.43 \\
                                     &AD2R&          1.98  &  \underl{1.81}&          1.98 &         39.96 & $>100$ & 39.88 \\
                                   
\hline
\multirow{3}{*}{saturated fat}       &AD  &  \underl{0.16} &          0.36 & \underl{0.16} &         0.17  &   2.66 &            0.17 \\
                                     &ADR &  \underl{0.08} &          0.14 & \underl{0.08} & \underl{0.08} &   1.72 & \underline{0.08} \\
                                     &AD2R&           3.07 &          3.76 & \underl{3.06} &         3.27  & $>100$ &            3.27 \\
 \hline
\multirow{3}{*}{monounsaturated fat} &AD  &  \underl{0.33} &          0.63 & \underl{0.33} &         0.62  &   5.17 &  0.62 \\
                                     &ADR &  \underl{0.11} &          0.22 & \underl{0.11} &         0.23  &   3.39 &  0.26 \\
                                     &AD2R&          4.42  &         12.78 & \underl{4.40} &        29.04  & $>100$ & 29.06 \\
 \hline
\multirow{3}{*}{polyunsaturated fat} &AD  &           0.57 & \underl{0.43} &          0.57 &          1.02 &  21.93 & 1.02 \\
                                     &ADR &  \underl{0.26} & \underl{0.26} & \underl{0.26} &          0.51 &  13.18 & 0.51 \\
                                     &AD2R&  \underl{3.52} &         10.56 & \underl{3.52} &         86.75 & $>100$ & 86.57  \\
\hline
\multirow{3}{*}{cholesterol}         &AD  &           0.32 &          0.35 & \underl{0.28} &               &  4.22  &  1.16 \\
                                     &ADR &           0.13 &          0.15 & \underl{0.12} &               &  2.45  &  0.51 \\
                                     &AD2R&           3.40 & \underl{2.98} &          3.34 &               & $>100$ & 11.01 \\
\hline
\end{tabular}
\end{center}
\label{ad13}
\end{table}
\normalsize

For the energy intake data, the estimate of the skewness parameter (${\color{black}{\lambda}}$) is indistinguishable from zero for the four models. Its estimates are close to zero, and the standard errors are relatively large. 
It suggests that log-symmetric models may be appropriate. Table \ref{energia2} presents the parameter estimates and the goodness-of-fit statistics for the log-t, log-normal, log-power exponential, and log-slash model fits. Comparing the figures in Tables \ref{energia1} and \ref{energia2} one may notice that AD and ADR, and the estimates for $\mu$, $\sigma$ and $\tau$ do not change much and the AICs are slightly smaller for the log-symmetric models. On the other hand, AD2R dropped from 112.25 (BCCG model) to 48.17 (log-normal model) and from 4.18 (BCPE model) to 2.70 (log-PE model). The change in AD2R is small when one moves from the BCT model to the log-t model or from the BCSlash model to the log-slash model.
Also, the quantile residuals plots (not shown) are similar to those presented in Figure \ref{residuos}. Taking parsimony into account, we may conclude that the best fitting model is the log-t model.

\begin{table}[H]
\normalsize \caption{Descriptive statistics, parameter estimates (standard error in parentheses) and goodness-of-fit measures;
animal protein intake data.}
\vspace{-0.8cm}
\begin{small}
\begin{center}
\begin{tabular}{l|c|c|c|c|c|c|c|c}
\hline
\hline
\multicolumn{3}{c}{\multirow{3}{*}{statistics (mg)}} \vline&  min & .25-quantile & median &  mean (s.d.) & .75-quantile & max \\
\cline{4-9}
\multicolumn{3}{c} {  }      \vline& 0.02 &        30.51 &  44.26 & 52.27  &        67.59 & 207.10  \\
\multicolumn{3}{c} {  }      \vline&      &              &        & (34.02)&              &          \\

\hline
\hline
distribution    & \multicolumn{2}{c}{BCT} \vline & \multicolumn{2}{c}{BCCG} \vline & \multicolumn{2}{c}{BCPE} \vline & \multicolumn{2}{c}{BCSlash} \\
\hline
$\mu$&\multicolumn{2}{c}{45.74 (1.46)} \vline &\multicolumn{2}{c}{46.37 (1.50)} \vline &\multicolumn{2}{c}{44.73 (1.36)} \vline & \multicolumn{2}{c}{45.83 (1.47)}   \\
$\sigma$&\multicolumn{2}{c}{0.55 (0.03)} \vline &\multicolumn{2}{c}{0.67 (0.02)} \vline &\multicolumn{2}{c}{0.70 (0.03)} \vline  & \multicolumn{2}{c}{0.45 (0.02)}   \\
${\color{black}{\lambda}}$  &\multicolumn{2}{c}{0.42 (0.07)} \vline  &\multicolumn{2}{c}{0.44 (0.06)} \vline &\multicolumn{2}{c}{0.42 (0.07)} \vline & \multicolumn{2}{c}{0.42 (0.10)}   \\
$\tau$ &\multicolumn{2}{c}{4.90 (0.92)} \vline  &\multicolumn{2}{c}{          } \vline &\multicolumn{2}{c}{1.24 (0.12)} \vline  & \multicolumn{2}{c}{  }\\
$q$    &\multicolumn{2}{c}{    }       \vline  &\multicolumn{2}{c}{          } \vline & \multicolumn{2}{c}{  }        \vline  & \multicolumn{2}{c}{3.02 (0.35)} \\
\hline
\hline
AIC & \multicolumn{2}{c}{3514.4} \vline & \multicolumn{2}{c}{3526.30} \vline & \multicolumn{2}{c}{3511.4} \vline & \multicolumn{2}{c}{3517.0} \\ 
AD  & \multicolumn{2}{c}{  0.35} \vline & \multicolumn{2}{c}{   1.46} \vline & \multicolumn{2}{c}{   0.51}\vline & \multicolumn{2}{c}{   0.43}\\
ADR & \multicolumn{2}{c}{  0.16} \vline & \multicolumn{2}{c}{   0.70} \vline & \multicolumn{2}{c}{   0.26}\vline & \multicolumn{2}{c}{   0.18} \\
AD2R& \multicolumn{2}{c}{  3.19} \vline & \multicolumn{2}{c}{  23.33} \vline & \multicolumn{2}{c}{   3.42}\vline & \multicolumn{2}{c}{   3.64} \\
\hline
\hline
\end{tabular}
\end{center}
\end{small}
\label{protanimal1}
\end{table}
\begin{table}[H]
\normalsize \caption{Descriptive statistics, parameter estimates (standard errors in parentheses) and goodness-of-fit measures;
energy intake data.}
\vspace{-0.8cm}
\begin{small}
\begin{center}
\begin{tabular}{l|c|c|c|c|c|c|c|c}
\hline
\hline
\multicolumn{3}{c}{\multirow{3}{*}{statistics (kcal)}} \vline&  min   & .25-quantile & median  &  mean (s.d.) & .75-quantile & max  \\
\cline{4-9}
 \multicolumn{3}{c}  { } \vline& 298.80 & 1356.00      & 1723.00 & 1868.00      & 2197.00      & 8370.00 \\
 \multicolumn{3}{c}  { } \vline&        &              &         & (838.35)     &              &         \\

\hline
\hline
distribution    & \multicolumn{2}{c}{BCT} \vline & \multicolumn{2}{c}{BCCG} \vline & \multicolumn{2}{c}{BCPE} \vline & \multicolumn{2}{c}{BCSlash}  \\
\hline
$\mu$   &\multicolumn{2}{c}{1725.00 (34.48)} \vline  &  \multicolumn{2}{c}{1726.00 (36.86)} \vline  & \multicolumn{2}{c}{1724.00 (34.14)} \vline  & \multicolumn{2}{c}{1725.00 (34.69)}   \\
$\sigma$&\multicolumn{2}{c}{0.34    (0.02)}  \vline  &  \multicolumn{2}{c}{0.41    (0.02)} \vline  & \multicolumn{2}{c}{0.41    (0.02)}\vline  & \multicolumn{2}{c}{0.29 (0.01)}    \\
${\color{black}{\lambda}}$  &\multicolumn{2}{c}{0.05    (0.12)} \vline   &  \multicolumn{2}{c}{0.07    (0.10)} \vline  & \multicolumn{2}{c}{0.06    (0.11)} \vline  & \multicolumn{2}{c}{0.05 (0.14)}    \\
$\tau$ &\multicolumn{2}{c}{6.14    (1.37)} \vline   &  \multicolumn{2}{c}{              } \vline  & \multicolumn{2}{c}{1.40    (0.14)}  \vline  & \multicolumn{2}{c}{ }   \\
$q$ &\multicolumn{2}{c}{} \vline   &  \multicolumn{2}{c}{              } \vline  & \multicolumn{2}{c}{}  \vline  & \multicolumn{2}{c}{3.71 (0.44)}   \\
\hline
\hline
AIC & \multicolumn{2}{c}{5861.9} \vline & \multicolumn{2}{c}{5876.1} \vline & \multicolumn{2}{c}{5864.0} \vline & \multicolumn{2}{c}{5863.3} \\
AD  & \multicolumn{2}{c}{   0.19}\vline & \multicolumn{2}{c}{   1.13}\vline & \multicolumn{2}{c}{   0.25}\vline & \multicolumn{2}{c}{   0.23}\\
ADR & \multicolumn{2}{c}{   0.10}\vline & \multicolumn{2}{c}{   0.58}\vline & \multicolumn{2}{c}{   0.14}\vline & \multicolumn{2}{c}{   0.12}\\
AD2R& \multicolumn{2}{c}{   1.78}\vline & \multicolumn{2}{c}{ 112.25}\vline & \multicolumn{2}{c}{   4.18}\vline & \multicolumn{2}{c}{   2.15}\\
\hline
\hline
\end{tabular}
\end{center}
\end{small}
\label{energia1}
\end{table}
\begin{table}[H]
\normalsize \caption{Parameter estimates (standard errors in parentheses) and goodness-of-fit statistics for the $\log$-$t$, $\log$-normal, $\log$-power exponential and $\log$-slash  models; energy intake data.}
\vspace{-0.8cm}
\begin{small}
\begin{center}
\begin{tabular}{l|c|c|c|c|c|c|c|c}
\hline
\hline
distribution    & \multicolumn{2}{c}{$\log$-t} \vline & \multicolumn{2}{c}{$\log$-normal} \vline & \multicolumn{2}{c}{$\log$-PE} \vline & \multicolumn{2}{c}{$\log$-slash}  \\
\hline
$\mu$   &    \multicolumn{2}{c}{1722.00 (34.44)} \vline  &  \multicolumn{2}{c}{1717.00 (36.74)} \vline  & \multicolumn{2}{c}{1721.00 (34.09)} \vline  & \multicolumn{2}{c}{1722.00 (34.63)}   \\
$\sigma$&    \multicolumn{2}{c}{0.34    (0.02)}  \vline  &  \multicolumn{2}{c}{0.41    (0.02)} \vline  & \multicolumn{2}{c}{0.41    (0.02)} \vline  & \multicolumn{2}{c}{0.28 (0.01)}    \\
$\tau$  &    \multicolumn{2}{c}{6.09    (1.34)} \vline   &  \multicolumn{2}{c}{              } \vline  & \multicolumn{2}{c}{1.40    (0.14)} \vline  & \multicolumn{2}{c}{}    \\
$q$  &    \multicolumn{2}{c}{} \vline   &  \multicolumn{2}{c}{              } \vline  & \multicolumn{2}{c}{} \vline  & \multicolumn{2}{c}{3.71 (0.44)}    \\
\hline
\hline
AIC & \multicolumn{2}{c}{5860.01} \vline & \multicolumn{2}{c}{5874.72} \vline & \multicolumn{2}{c}{5862.27}\vline & \multicolumn{2}{c}{5861.46} \\
AD  & \multicolumn{2}{c}{   0.19} \vline & \multicolumn{2}{c}{   1.16} \vline & \multicolumn{2}{c}{   0.26} \vline & \multicolumn{2}{c}{  0.23}\\
ADR & \multicolumn{2}{c}{   0.10} \vline & \multicolumn{2}{c}{   0.55} \vline & \multicolumn{2}{c}{   0.15} \vline & \multicolumn{2}{c}{  0.12}\\
AD2R& \multicolumn{2}{c}{   1.80} \vline & \multicolumn{2}{c}{  48.17} \vline & \multicolumn{2}{c}{   2.70} \vline & \multicolumn{2}{c}{  2.11}\\
\hline
\hline
\end{tabular}
\end{center}
\end{small}
\label{energia2}
\end{table}

{\color{black}{For the BCT model fitted to the animal protein intake data the estimates of $\mu$, $\sigma$, $\lambda$, and $\tau$ reported in Table \ref{protanimal1} are $\widehat \mu_P=45.74$, $\widehat \sigma_P=0.55$, $\widehat \lambda_P=0.42$, and  $\widehat\tau_P=4.90$, respectively. For the log-t model fitted to the energy intake data the estimates of $\mu$, $\sigma$, and $\tau$ are $\widehat \mu_E=1722.00$, $\widehat \sigma_E=0.34$, and $\widehat \tau_E=6.09$, respectively; see Table \ref{energia2}. Because $\widehat \sigma_P  \widehat \lambda_P$ is small, $\widehat \mu_P$ may be seen as an estimate of the population median of the animal protein intake. As expected, $\widehat \mu_P$ is close to the sample median (44.26). Similarly, $\widehat \mu_E$, which is an estimate of the population median of the energy intake, is close to the sample median (1723.00). Additionally, $\widehat \sigma_P$ is considerably larger than $\widehat \sigma_E$ indicating that the relative dispersion of the population intake of  animal protein is larger than that of energy. The estimates of the degrees of freedom parameter $\tau$, both not large, reveal the need for right heavy tailed distributions for an adequate fit for the data.}}


\section{Concluding remarks}\label{conc}

This paper proposed a new class of distributions, the Box-Cox symmetric distributions. It contains some well known
distributions as special cases and allows the definition of new distributions, such as the Box-Cox slash distribution.
It is particularly suitable for inference on positively skewed, possibly heavy-tailed, data.
It permits easy parameter interpretation, a desirable feature for modeling.

There is clear possibility for extension to regression models. Some or all the parameters of the BCS
distributions may be modeled by a link function and a linear or nonlinear regression model structure. The GAMLSS
framework \citep{Rigby} is a natural tool for implementing BCS regression models. It allows the regression structure to
include parametric and nonparametric terms and random effects. Box-Cox t, Box-Cox Cole-Green, and Box-Cox power exponential models
are already implemented in {\tt gamlss} package in {\tt R}.

Some BCS distributions include an extra parameter; e.g., the degrees of freedom parameter of the BCT distribution.
We have not faced convergence problems or unrealistic estimation when the additional parameter is estimated simultaneously with the
others. It should be noticed that the sample sizes in our applications were relatively large ($n=368$). In small samples, it may be
advisable to set a grid of values for the extra parameter and choose the value that provides the best fit according to
the chosen criteria.

Applications to data on intake of several nutrients illustrated that the BCS distributions are useful in practice.
The data correspond to the first 24-hour dietary recall interview for the individuals in the sample.
It is part of our current research to develop Box-Cox symmetric models with random and mixed effects
to model nutrients intake data taken from repeated 24-hour recalls.

{\color{black}{As a final remark, we recall that a comprehensive study on the right tail heaviness of the Box-Cox 
symmetric distributions was presented. For future possible investigation, it might be interesting to search for
skewness-kurtosis boundaries allowing the existence of BCS distributions, as in \citet{Jondeau}.}}

\section*{Acknowledgments}

We thank Jos\'e Eduardo Corrente for providing the data used in this study, and Eliane C. Pinheiro for helpful discussions. We acknowledge the financial support of FAPESP, CAPES and CNPq (Brazil). We are grateful to the associate editor and two anonymous referees for constructive comments and suggestions.

{\color{black}{
\section*{Appendix}

In this appendix we give the first and second derivatives of the log-likelihood function with respect to the parameters. 
Let $z=h(y;\mu,\sigma,\lambda)$, where $h(y;\mu,\sigma,\lambda)$ is given in (\ref{ztrans}), $\varpi=-2r'(z^2)/r(z^2)$,
and 
$\xi=r((\sigma \lambda)^{-2}) / R((\sigma |\lambda|)^{-1}).$ 
We have
$$\displaystyle \frac{\partial z}{\partial \mu}=-\frac{1}{\mu \sigma} \left(\frac{y}{\mu}\right)^\lambda \xrightarrow[\lambda \to  0]{} -\frac{1}{\mu \sigma},$$  \ \
$$\displaystyle \frac{\partial z}{\partial \lambda}=\frac{1}{\sigma \lambda^2} \left\{ 1+ \left(\frac{y}{\mu}\right)^{\lambda} \left[-1    + \lambda \log\left(\frac{y}{\mu}\right) \right] \right\} \xrightarrow[\lambda \to  0]{} \frac{1}{2 \sigma}\left[ \log\left(\frac{y}{\mu}\right)\right]^2,$$ 
$$\displaystyle \frac{\partial^2 z}{\partial \mu^2}= \frac{(\lambda+1)}{\mu^2 \sigma} \left(\frac{y}{\mu}\right)^\lambda \xrightarrow[\lambda \to  0]{} \frac{1}{\mu^2 \sigma},$$ \ \
  $$\displaystyle \frac{\partial^2 z}{\partial \lambda^2}= \frac{1}{\sigma \lambda^3}\left\{-2+ \left(\frac{y}{\mu}\right)^{\lambda} \left[2- 2 \lambda \log \left(\frac{y}{\mu}\right)+\lambda^2 \left(\log\left(\frac{y}{\mu}\right)\right)^2 \right] \right\} \xrightarrow[\lambda \to  0]{} \frac{1}{3 \sigma} \left[\log\left(\frac{y}{\mu} \right)\right]^3,$$ 
$$
  \displaystyle \frac{\partial^2 z}{\partial \mu\partial\lambda}= -\frac{1}{\mu \sigma} \left(\frac{y}{\mu}\right)^\lambda \log\left(\frac{y}{\mu}\right)  \xrightarrow[\lambda \to  0]{} -\frac{1}{\mu \sigma}\log\left(\frac{y}{\mu}\right) .\ \
$$


\hspace{0.5cm}

Let $\ell$ denote the log-likelihood for a single observation $y$. We have
$$\ell= (\lambda-1) \log y -  \lambda \log \mu - \log \sigma 
+\log r(z^2)- \log R\left(\frac{1}{\sigma |\lambda|}\right), \nonumber $$
if $\lambda \neq 0$; the last term in $\ell$ is zero if $\lambda=0$.
The first derivatives of $\ell$ are given by
$$
\frac{\partial \ell}{\partial \mu}= -\frac{\lambda}{\mu}-\varpi z\frac{\partial z}{\partial \mu},
$$
$$
\displaystyle \frac{\partial \ell}{\partial \sigma}=\left\{
\begin{array}{ll}
\displaystyle {-\frac{1}{\sigma}+ \frac{\varpi z^2}{\sigma}+ \frac{\xi}{\sigma^2 |\lambda|}}, & {\mbox{if} \quad \lambda\neq0}, \\
\displaystyle {-\frac{1}{\sigma}+ \frac{\varpi z^2}{\sigma}}, & { \mbox{if} \quad \lambda= 0,}
\end{array}
\right.
$$
$$
	\frac{\partial \ell}{\partial \lambda}= \log\left(\frac{y}{\mu}\right)-\varpi
	z\frac{\partial z}{\partial \lambda}+ \mbox{sign}(\lambda)\frac{ \xi}{\sigma \lambda^2}. 
$$
The second derivatives of $\ell$ are given by
$$
\frac{\partial^2 \ell}{\partial \mu^2}= \frac{\lambda}{\mu^2}-\left( z \frac{d \varpi}{d z} +\varpi \right) \left(\frac{\partial z}{\partial \mu}\right)^2-\varpi z \frac{\partial^2 z}{\partial \mu^2}, 
$$
$$
\displaystyle \frac{\partial^2 \ell}{\partial \sigma^2}=\left\{
\begin{array}{ll}
\displaystyle {\frac{1}{\sigma^2}- \frac{z^3}{\sigma^2} \frac{d \varpi}{d z}- \frac{3 \varpi z^2}{\sigma^2}+\frac{1}{\sigma^2 |\lambda|} \frac{\partial \xi}{\partial \sigma}-\frac{2 \xi}{\sigma^3 |\lambda|}}, & {\mbox{if} \quad \lambda\neq0}, \\
\displaystyle {\frac{1}{\sigma^2}- \frac{z^3}{\sigma^2} \frac{d \varpi}{d z}- \frac{3 \varpi z^2}{\sigma^2}}, & { \mbox{if} \quad \lambda= 0,}
\end{array}
\right.
$$
$$
\frac{\partial^2 \ell}{\partial \lambda^2}= -\left( z \frac{d \varpi}{d z} +\varpi \right) \left(\frac{\partial z}{\partial \lambda}\right)^2-\varpi z \frac{\partial^2 z}{\partial \lambda^2}+\mbox{sign}(\lambda)\left(\frac{1}{\sigma \lambda^2} \frac{\partial \xi}{\partial \lambda}-\frac{2 \xi}{\sigma \lambda^3}\right),
$$
$$
\frac{\partial^2 \ell}{\partial \mu\partial\sigma}=\frac{z}{\sigma} \frac{\partial z}{\partial \mu} \left( z \frac{d \varpi}{d z} + 2 \varpi \right),
$$
$$
\frac{\partial^2 \ell}{\partial \mu \partial\lambda}= -\frac{1}{\mu}-\left( z \frac{d \varpi}{d z} +\varpi \right) \frac{\partial z}{\partial \mu}\frac{\partial z}{\partial \lambda}-\varpi z \frac{\partial^2 z}{\partial \mu \partial\lambda},
$$
$$
\frac{\partial^2 \ell}{\partial \sigma \partial\lambda}= \frac{z}{\sigma} \frac{\partial z}{\partial \lambda} \left( z \frac{d \varpi}{ d z}+ 2 \varpi \right)+ \frac{1}{ \sigma^2 |\lambda|} \frac{\partial \xi}{\partial \lambda} - \mbox{sign}(\lambda) \frac{\xi}{\sigma^2 \lambda^2}.
$$
The first and second derivatives of $\ell$ are obtained after plugging the derivatives of $z$ given above. 

Note that the first derivatives of $\ell$ depend on the weighting function $\varpi$ ($\varpi$ is given in Table \ref{funcaopeso} for some distributions). Consequently, $d\varpi/d z$ appears in all the second derivatives of $\ell$. Note that $\partial \ell/\partial \sigma$ and $\partial \ell/\partial \lambda$  involve $\xi$, which in turn depends on the particular distribution in the BCS class and the truncation set. The first derivatives of $\xi$ appear in $\partial^2 \ell/\partial \sigma^2$, $\partial^2 \ell/\partial \lambda^2$ and $\partial^2 \ell/\partial \sigma\partial\lambda$. The stability of the terms that involve $\xi$ and its first derivatives around $\lambda=0$ may vary according to different distributions. For instance, they may be unstable for the Box-Cox t distribution with small degrees of freedom parameter. Yet, a simulation study of the type I error probability of the likelihood ratio test of ${\rm H}_0: \lambda=0$ in the Box-Cox t model for different values of the degrees of freedom parameter performed well; see Section \ref{inference}.
}}

\bibliographystyle{dcu}
\bibliography{referencias}

\end{document}